\documentclass[a4paper,12pt]{article}
\usepackage{epsfig}
\usepackage{times}

\begin{document}

\title{\Large\bf Nonmodal energy growth and optimal perturbations 
in compressible plane Couette flow\footnote{Physics of Fluids, vol. {\bf 18}, 034103 (2006, March)}
}
\author{M. Malik$^{1}$, Meheboob Alam$^{2}$\footnote{Corresponding Author.
Email: meheboob@jncasr.ac.in} and J. Dey$^{1}$ \\
$^{1}$Department of Aerospace Engineering, 
 Indian Institute of Science,\\ Bangalore 560012, India\\
$^{2}$Engineering Mechanics Unit, Jawaharlal Nehru Center for Advanced\\
Scientific Research, Jakkur P.O., Bangalore 560064, India
}

\date{\today}
                                                                                              
\maketitle

\begin{abstract}
Nonmodal transient growth studies and estimation of optimal 
perturbations have been made for the compressible plane Couette flow 
with three-dimensional disturbances. 
The steady mean flow is characterized by a  non-uniform shear-rate and a
varying temperature across the wall-normal direction 
for an appropriate  perfect gas model.
The maximum amplification of perturbation energy  over time,
$G_{\max}$, is found to increase with increasing Reynolds number ${\it Re}$,
but decreases with increasing Mach number $M$.
More specifically, the optimal energy amplification $G_{\rm opt}$ 
(the supremum of $G_{\max}$ over both the streamwise and spanwise wavenumbers)
is maximum in the incompressible limit and decreases monotonically as $M$ increases.
The corresponding optimal streamwise wavenumber,
$\alpha_{\rm opt}$, is non-zero at $M=0$, increases with
increasing $M$, reaching a maximum for some value of $M$
and then decreases, eventually becoming zero at high Mach numbers.
While the pure streamwise vortices are the optimal patterns 
at high Mach numbers (in contrast to incompressible Couette flow), 
the modulated streamwise vortices
are the optimal patterns for low-to-moderate values of the Mach number.
Unlike in incompressible shear flows,
the streamwise-independent modes in the present flow do not follow
the scaling law $G(t/{\it Re}) \sim {\it Re}^2$,
the reasons for which are shown to be tied to the dominance of some terms
(related to density and temperature fluctuations) in the linear stability operator. 
Based on a detailed nonmodal energy analysis,
we show  that the  transient energy growth occurs due to
the transfer of energy from the mean flow to perturbations via an inviscid 
{\it algebraic} instability.
The decrease of transient growth with increasing Mach number is also shown to be tied to 
the decrease in the energy transferred from the mean flow ($\dot{\mathcal E}_1$)
in the same limit.
The sharp decay of the viscous eigenfunctions with increasing  Mach number
is responsible for the decrease of $\dot{\mathcal E}_1$
for the present mean flow.
\end{abstract}

\hrulefill

\section{Introduction}

The linear stability (LS) analysis of compressible flows is of 
interest from the viewpoint of basic understanding on transition scenarios 
in such fluids~\cite{LL46,Mack69,Mack84}, and 
also because of its relevance in many high speed aerodynamic design problems. 
Since the growth of a small disturbance can lead to the flow breakdown, 
an understanding of different growth mechanisms becomes more relevant. 
The normal mode approach, where the initial value problem is reduced to an 
eigenvalue problem by considering the disturbance growth/decay in terms of 
an exponential time dependence, has been widely studied in the past.
For more than a decade, however, the traditional modal stability analysis 
has been complemented by the analysis of transiently growing 
perturbations for  various shear 
flow configurations~\cite{Orr07,BF92,RH93,TTRD93,WAL95}.
Such transient growth analyses have revealed that a flow can sustain
large transient growth of energy
in the parameter space that is stable according to the LS theory.
It has been shown that the transient growth analysis can provide
a possible reason for the experimentally observed critical Reynolds number 
(${\it Re}_{\rm cr}$) of various canonical flow configurations
being less than that predicted by the LS theory
(for a detailed review see the recent book of Schmid and Henningson~\cite{SH01}).

It is well known that there are flow configurations 
(e.g. the plane Couette flow, pipe flow, etc.) that are stable according to the LS theory, 
but have been shown to have a finite ${\it Re}_{\rm cr}$ in  experiments.
Therefore the nonlinear effects may destabilize these flows 
in the stable parameter region of the LS theory. 
However, for the nonlinearities in the governing equations to take over, 
the amplitudes of the disturbances must be large. 
Therefore, there may be a linear mechanism that causes an infinitesimally small
disturbance already present in the flow to grow initially, so that the nonlinearities 
could act upon them. Such a mechanism for the transient growth exists
due to the nonmodal growth caused by the non-orthogonality of the LS operator~\cite{SH01}.
In the stable region of the parameter space, even though each eigenstate decays 
during its evolution in space or time, a superposition 
of such eigenstates has potential for the transient inviscid growth before being
stabilized by the viscosity at the rate of the least decaying mode. 
The experimental confirmation of this transition scenario
comes from the observation that {\it streaks} precede 
the breakdown of laminar shear flows. Since the transient  energy growth is maximum 
for {\it streamwise vortices} which subsequently give birth to
streaks, it suggests that this  transient growth mechanism
provides a robust route to flow breakdown via disturbances
that are stable according to LS theory.

Even though the  transient growth analysis has been extensively
studied for incompressible flows~\cite{BF92,Gust91,RH93,TTRD93,WAL95,SH01,KL00},
similar studies on compressible flows are scarce~\cite{HSH96,HH98,FI00,TR03}.
For compressible boundary layers, Hanifi, Schmid and Henningson~\cite{HSH96} 
have found the maximum transient energy growth ($G_{max}$)
increases with both Reynolds number ($Re$) and Mach number ($M$).
They have used an energy norm such 
that the pressure-related energy transfer rate is equal to zero. 
They also cofirmed that the streamwise vortices 
are the optimal patterns as in incompressible flows.
Farrell and Ioannou~\cite{FI00} have investigated the transient energy growth 
of 2D perturbations in the Couette flow (uniform shear flow) of a polytropic fluid
with constant viscosity coefficients. 
They found that $G_{max}$  increases with Mach number.
Recently, Tumin and Reshotko~\cite{TR03} have developed
a `spatial' model of transient growth phenomena in 
compressible boundary layers.

The present work deals with the `temporal'  stability analyses
of the compressible plane Couette flow.
The modal linear stability analysis of this flow has been reported 
by many investigators in the past~\cite{Glat89,DEH94,HZ98}.
In the incompressible limit this flow remains stable
but instability is possible for a range of supersonic Mach numbers,
the origin of which is tied to a family of acoustic modes~\cite{DEH94,HZ98}. 
Duck, Erlebacher and Hussaini~\cite{DEH94} have classified these acoustic modes
into two distinct families depending on their phase speeds 
(see Fig. 2, the detailed description of this figure is provided in Sec. 3.1):
the {\it odd-modes} (I, III, ...) that have phase speeds greater than unity 
in the limit of zero streamwise wavenumber ($\alpha$),
and the {\it even-modes} (II, IV, ...)
that have phase speeds  less than zero  as $\alpha\to 0$.
(For the plane Couette flow, 
the non-dimensional velocities of the walls are bounded between $0$ and $1$
since the top-wall velocity has been used for non-dimensionalization, see Fig. 1.)
From an asymptotic analysis of the inviscid equations,
they showed that the mode-I remains neutrally stable,
but the mode-II can become  unstable in the inviscid limit for large $\alpha$.
Hu and Zhong~\cite{HZ98} studied the same viscous plane Couette flow
problem numerically, and reported new viscous instabilities.
In particular, they showed  that the mode-I can become unstable 
at finite Reynolds numbers due to
the effects of viscosity  but the ranges of Mach numbers and wavenumbers are very narrow.
They further showed that the viscosity plays a dual role of stabilizing/destabilizing
the mode-II instability.
The inviscid mode-II instabilty of Duck et al. that occurs at large $\alpha$
is weakly stabilizied by viscosity, but the same mode-II is also unstable for
low streamwise wavenumbers purely due to  viscous effects~\cite{HZ98}.
Even though the higher-order even modes (IV,...) can also become unstable,
the mode-II instablity represents the dominant instability for the 
compressible plane Couette flow.

In this paper, the  nonmodal transient energy growth analysis is reported 
for the compressible plane Couette flow of an appropriate
perfect gas model for  three-dimensional disturbances.
The governing equations and the mean flow are detailed in Sec. 2.
The linear stability problem is formulated in Sec. 3, 
and the related numerical method and its validation are detailed in Sec. 3.1.
For the tranisent growth analysis, the disturbance size is measured
in terms of the Mack energy-norm~\cite{Mack69,Mack84}.
The results on the energy growth, the optimal perturbations and their structure,
and the scalings of maximum transient growth are presented in Sec. 4.
The constituent energies transferred from the mean flow, 
the viscous dissipation, the thermal diffusion 
and the shear-work during the transient growth are  investigated in Sec. 5 
for the initial perturbation configuration that would reach 
the optimal configuration at a later time.
The inviscid limit of stability equations are analysed in Sec. 5.1. 
The conclusions are provided in Sec. 6.

\section{Governing Equations}

We consider the plane Couette flow of a perfect gas
of density $\rho^*$ and temperature $T^*$, driven by the relative motion
of two parallel walls that are separated by a distance $h^*$.
The top wall moves with a velocity $U_1^*$ and the lower wall is stationary,
with the top-wall temperature being held fixed at $T_1^*$.
Note that the quantities with a superscript $*$ are dimensional,
and the subscript $1$ refers to the quantities at the top wall.
The velocity field is denoted by ${\bf u}^*(x^*,y^*,z^*,t^*) = (u^*, v^*, w^*)^T$,
with $u^*$, $v^*$ and $w^*$ being the velocity components in the
streamwise ($x^*$), wall-normal ($y^*$) and spanwise ($z^*$) directions, respectively.
For non-dimensionalization, we use the separation between the two walls $h^*$ as
the length scale, the top wall velocity, $U_1^*$, and temperature, $T_1^*$,
 as the velocity and temperature scale, respectively,
and the inverse of the overall shear-rate, $U_1^*/h^*$, as the time scale.
The governing equations are the three momentum equations, 
continuity, energy and state equations which are omitted for
the sake of brevity.
The flow is described by the Reynolds number $Re$,
the Prandtl number $\sigma$ and the Mach number $M$, defined via
\begin{equation}
{\it Re} = U^*_1\rho^*_1h^*/\mu^*_1, \quad  \ \sigma = \mu^*c_p/K^*, \quad
  \ \mbox{and} \quad \  M=U_1^*/\sqrt{\gamma R T_1^*}.
\end{equation}
Here $\gamma=c_p^*/c_v^*$ is the ratio of specific heats, 
$K^*$  the thermal conductivity, and $R=c_p^*-c_v^*$  the universal gas constant; 
the Prandtl number is assumed to be a constant, $\sigma=0.72$
and $\gamma=1.4$.  
For the shear viscosity $\mu$, we use the Sutherland formula
\begin{equation}
  \mu = \frac{T^{3/2}(1+C)}{(T+C)}, 
 \quad \mbox{with} \quad C=0.5.
\label{eqn_viscosity}
\end{equation}
Following Stokes' assumption, we set the bulk viscosity to zero
(i.e., $\zeta=0$) such that
\[
    \lambda = \zeta - \frac{2}{3}\mu = - \frac{2}{3}\mu.  
\]
To make a direct comparison with earlier linear stability results 
of the compressible plane Couette flow~\cite{DEH94,HZ98},
we have used the same constitutive model
for the transport coefficients of a perfect gas.

\subsection{Mean flow}
\label{meanflowseclab}

\begin{figure}[p]
\centerline{\psfig{figure=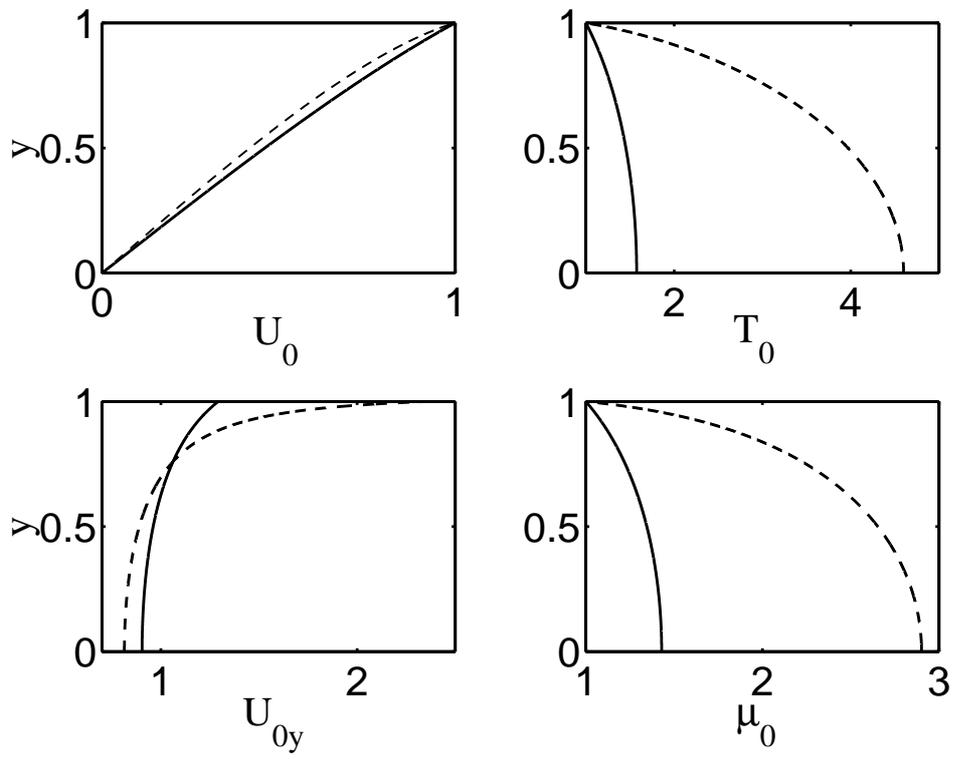,width=5in}}
\caption{Profiles of the mean velocity $U_0$,  temperature $T_0$,  
shear-rate $U_{0y}$  and viscosity $\mu_0$ for $M = 2$ (---$\!$---) and  $M = 5$ (-- -- --).
}
\label{fig:meanflow}
\end{figure}

The mean flow is that of the steady, fully developed plane
Couette flow for which the mean fields are given by
\begin{equation}
  [u,v,w] = [U_0(y), 0, 0],
\quad
 \rho = \rho_0(y),
\quad
 T = T_0(y).
\end{equation}
Hereafter the subscript $0$ is used to refer to the mean flow quantities.
The continuity and the $z$-momentum equations are
identically satisfied for this mean flow. From the $y$-momentum equation,
it is straightforward to verify that the mean pressure, $p_0$,
is constant, chosen to be $p_0=1$. Hence, from the
equation of state, the mean density is  related
to the mean temperature via $\rho_0(y) = 1/T_0(y)$.
The remaining $x$-momentum and energy equations for the mean flow are
\begin{eqnarray}
  \frac{\rm d}{{\rm d}y}\left(\mu_0\frac{{\rm d}U_0}{{\rm d}y}\right) &=& 0, \\
  \frac{\rm d}{{\rm d}y}\left(\frac{\mu_0}{\sigma}\frac{{\rm d}T_0}{{\rm d}y}\right) 
   + (\gamma -1)M^2\mu_0\left(\frac{{\rm d}U_0}{{\rm d}y}\right)^2 &=& 0,
\end{eqnarray} 
which have to be solved, satisfying the following boundary conditions:
\begin{equation}
  U_0(0)=0,
    \quad U_0(1)=1, 
    \quad T_0(0)=T_w, 
    \quad  T_0(1)=1,
\end{equation} 
where $T_w$ is the non-dimensional temperature of the lower wall. 
The energy equation can be solved exactly to yield the temparature 
in terms of the flow velocity~\cite{DEH94}:
\begin{equation}
   T_0 = T_r\left[r+(1-r)U_0-\left(1-\frac{1}{T_r}\right)U_{0}^{2}\right],
\end{equation}
where 
\begin{equation}
  T_r = 1+[(\gamma-1)\sigma M^2]/2
\end{equation}
is the recovery temperature, and $r = T_w/T_r$ is the temperature ratio. 
The streamwise  velocity, $U_0(y)$, is then calculated numerically 
by solving the following equation:
\begin{equation}
  \frac{{\rm d}U_0}{{\rm d}y} = \frac{\tau}{\mu(U_0)},
\end{equation}
using a fourth-order Runge-Kutta method.
The shear stress, $\tau$, is a constant
that must be determined iteratively such that $U_0(y)$ satisfies its
boundary values:
\begin{equation}
  U_0(0) =0 
  \quad
  \mbox{and}
  \quad
  U_0(1) =1.
\end{equation}
With $r = 1$, which corresponds to an adiabatic lower wall, 
the calculated profiles of the mean velocity ($U_0$),  
the temperature ($T_0$), the shear rate ($U_{0y}$) and the viscosity ($\mu_0$)
are shown in Fig.~\ref{fig:meanflow} 
for two representative values of the Mach number, $M=2$ and $5$.
(Note that the mean flow profiles do not depend on Reynolds number.)
Clearly, the shear-rate, the temperature and the viscosity are non-uniform along the
wall-normal direction; the deviation from the corresponding
uniform shear flow increases with increasing $M$.
We shall return back to discuss the effects of non-uniform mean flow on
certain aspects of the transient growth  results in Sec. 5.1.

%-------------------------
%-------------------------

\section{Linear Perturbation System}

We impose three-dimensional (3D) perturbations on the mean flow  described in Sec. ~\ref
{meanflowseclab}.
The flow quantities with small-amplitude disturbances 
are considered as $\bf q = q_0 + \hat{q}$, where $\bf q_0$ represents any mean flow 
quantity, and $\bf \hat{q}$ is its perturbation. For ${\bf q_0} \gg {\bf \hat
{q}}$, the governing equations are linearised with respect 
to ${\bf \hat{q}}$. Then the solutions of the linearized equations are 
assumed in the form of normal modes, i.e.,
\begin{eqnarray}
  {\bf \hat{q}}(x,y,z,t) &=& {\bf \tilde{q}}(y,t) \ \exp \ [i(\alpha x + \beta z)], 
  \label{modalxzdep}\\
  {\bf \tilde{q}}(y,t) & = & {\bf q'}(y) \ \exp \ (-i \omega t) 
  \label{modaltdep},
\end{eqnarray}
where $\alpha$ and $\beta$ are the streamwise and spanwise wavenumbers, respectively, 
and $\omega=\omega_r + i\omega_i$ is the complex frequency. 
The imaginary part of the complex frequency, $\omega_i$, represents the
growth/decay rate of perturbations, and its real part,  $\omega_r$,
is related to the phase speed of the perturbation via $c_r=\omega_r/\alpha$.
The flow is said to be asymptotically stable if $\omega_i<0$, unstable if $\omega_i>0$ and 
neutrally stable if $\omega_i=0$. Later for the 
transient growth analysis (Sec. 4), a summation over all $\omega$'s with appropriate 
coefficients will be implied on the right-hand-side of equation~(\ref{modaltdep}).

The perturbation pressure is removed from the linearized equations by using the 
perturbed equation of state
\begin {equation}
   \hat{p} = T_0 \hat\rho + \hat{T}/T_0.
\label{stateqnpert}
\end{equation}
Similarly, the perturbation viscosity is written in terms of perturbation temperature:
\begin{displaymath}
  \hat\mu = \mu_T \hat{T},
\end {displaymath}
where $\mu_T=({\rm d}\mu/{\rm d}T)_0$ is evaluated at mean flow conditions. 
Finally, the reduced ordinary differential eigensystem in five unknowns can  
be written as
\begin{equation}
 {\cal L} {\bf q'} = \omega {\rm I}{\bf q'},
\label{eigeqn}
\end {equation}
where ${\bf q'} = \{u',v',w',\rho',T'\}^{\bf \rm T}$ is the eigenfunction and 
${\rm I}$ the identity matrix. The elements of the linear operator
{$\cal L$} are given in the appendix. The boundary conditions are: 
\begin{equation}
 u'(0) = u'(1) = 0; \  v'(0) = v'(1) =0; \ w'(0) = w'(1) = 0; 
  T'(1) = \frac{{\rm d}T'}{{\rm d}y}(0) = 0.
\label{eigeqn_bc}
\end{equation}
Note that the temperature boundary condition corresponds to an isothermal upper wall and 
an adiabatic lower wall as in the work of Duck et al.~\cite{DEH94} and
Hu and Zhong~\cite{HZ98}.

\subsection{Numerical method and code validation}

The differential eigenvalue problem (\ref{eigeqn})-(\ref{eigeqn_bc}) is 
discretized by using the Chebyshev spectral method~\cite{Malik90}.
The transformation $\xi = (2y-1)$ is used  to 
map the physical domain $y \in [0,1]$ to the Chebyshev domain $\xi \in [-1,1]$. 
The equations are collocated at the Gauss-Lobatto collocation points:
\begin{equation}
  \xi_j = \cos\left(\frac{j\pi}{N}\right),
\quad
  j = 0,1,2,\ldots,N,
\end{equation}
which are the extrema of the $N$-order Chebyshev
polynomial $T_N$.
There are two sub-methods of 
the Chebyshev spectral method to discretize this eigenvalue problem (\ref{eigeqn}). 
In one method, any function $\phi$, say, is approximated by using a Chebyshev polynomial
of degree $N$:
\begin{equation}
   \phi(\xi) = \sum_{n=0}^{N} a_n {\rm T}_n(\xi),
\end{equation}
with the Chebyshev coefficients $a_n$ being treated as `unknowns'.
The derivatives of $\phi(\xi)$ are also found in terms of these polynomials
through a recurrence relation that they satisfy~\cite{SH01}. 
In the other method, 
the function $\phi$ is expressed in terms of its values
at the collocation points using an interpolant that uses Chebyshev polynomials,
and the derivatives are found from an 
interpolation formula
(see Malik~\cite{Malik90} for details). This latter 
method is convenient since it results in a simple eigensystem, 
whereas the former results in a generalized eigensystem. 
The $x$, $y$, $z$-momentum equations and the energy equation are replaced by the boundary 
conditions for $u'$, $v'$, $w'$ and $T'$ at walls, respectively. 
The unused $y$-momentum equation is used to replace the continuity equation at the walls  
(also termed as artificial boundary condition~\cite{Malik90}). 
Since the boundary conditions 
are independent of the eigenvalue $\omega$, the right hand side of the eqn~(\ref{eigeqn}) 
would be singular, as the rows corresponding to boundary points would be zero.
This singularity is removed  
by the row/column operations, resulting in a reduced system. 
The QR-algorithm of the Matlab software is used to solve this eigenvalue system.

\begin{figure}[p]
\begin{center}
\begin{tabular}{lr}
\begin{minipage}[t]{3.0in}
\begin{picture}(3.0,2.5)
\centerline{\psfig{figure=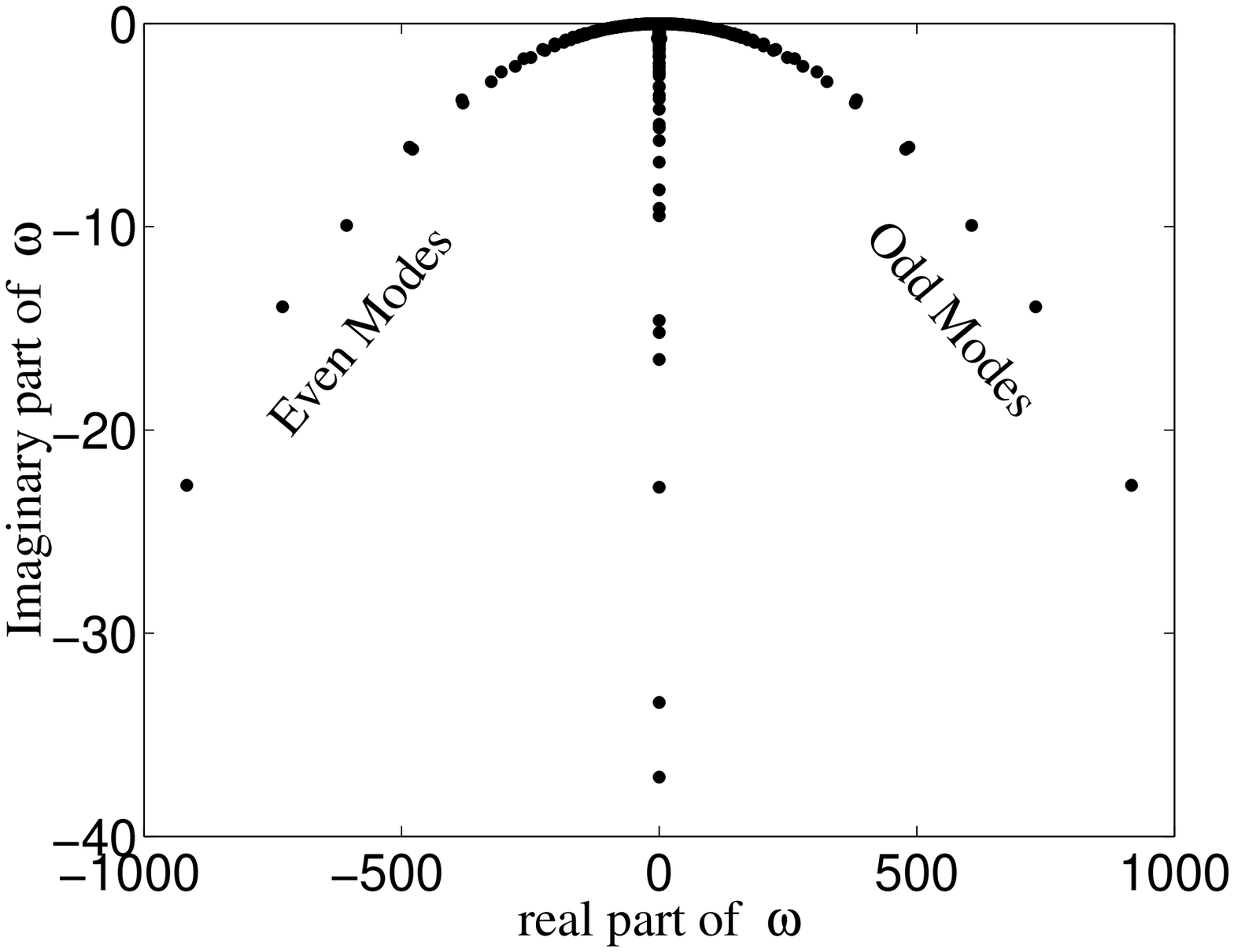,width=3.0in}}
\put(-3.1,2.1){(a)}
\end{picture}
\end{minipage}
\\ 
\begin{minipage}[t]{3.0in}
\begin{picture}(3.0,2.5)
\centerline{\psfig{figure=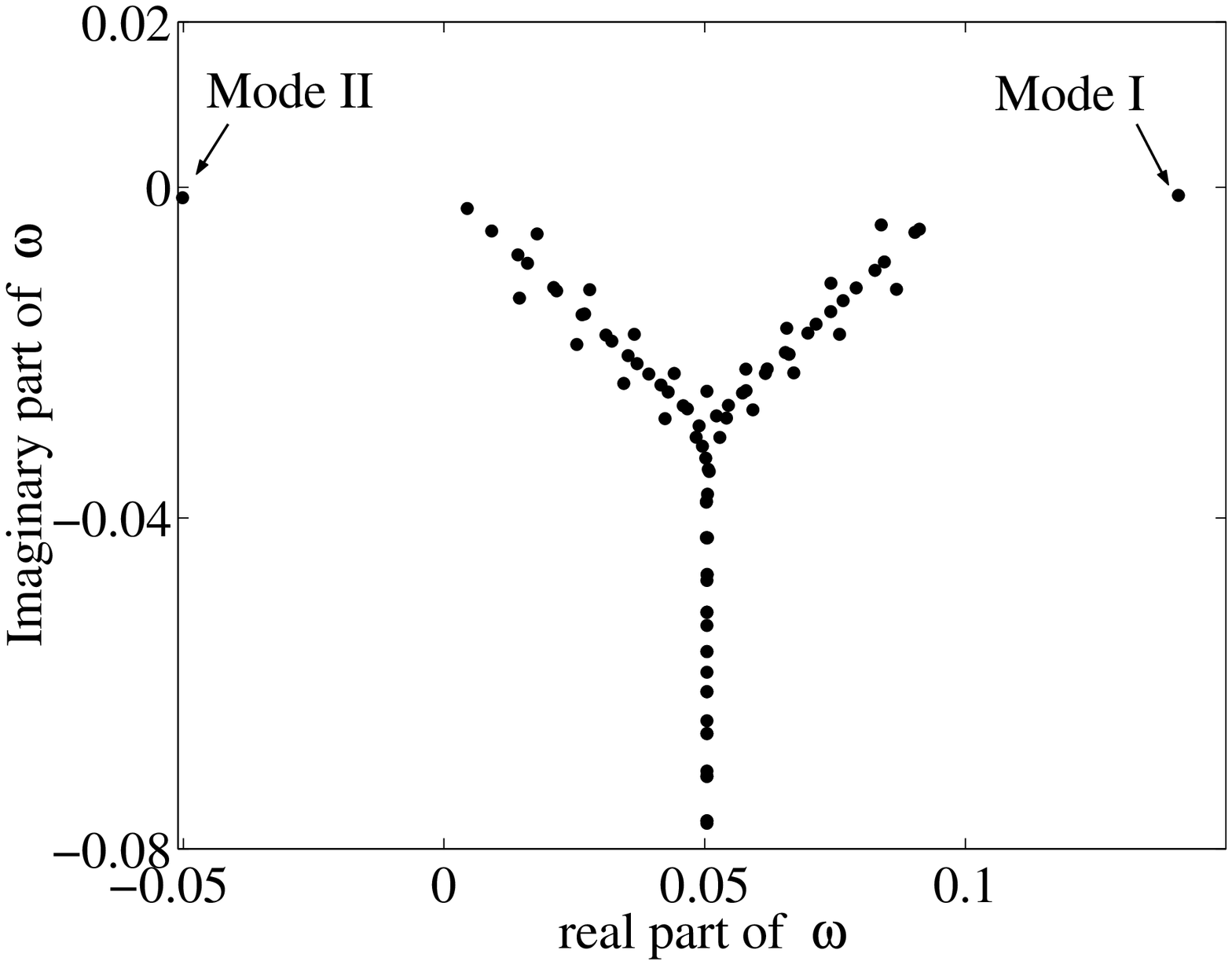,width=3.0in}}
\put(-3.1,2.1){(b)}
\end{picture}
\end{minipage}
\end{tabular}
\end{center}
\caption{Spectra for $M = 2$ and ${\it Re} = 2\times 10^5$:
$\alpha = 0.1 =\beta$; $N=100$.
Panel $b$ is the zoom of panel $a$.
}
\label{fig:fig_spectra1}
\end{figure}

To validate the stability code, we have compared our numerical results
on asymptotic stability with the results of Hu and Zhong~\cite{HZ98}.
For example, for the parameter values of $Re=2\times 10^5$, $M=2$, $\alpha=0.1$
and $\beta=0$ with $N=100$ collocation points, 
our data on both growth rate and frequency
match those of Hu and Zhong upto the fourth decimal place.
For this parameter set, the distribution of eigenvalues, $c = \omega/\alpha$,
is shown in Fig.~\ref{fig:fig_spectra1}($a$),
and the enlargement of Fig.~\ref{fig:fig_spectra1}($a$) is shown in
Fig.~\ref{fig:fig_spectra1}($b$) that portrays the well-known `Y'-branch of the spectra
that belongs to viscous modes.
Following the classification of Duck et al.~\cite{DEH94},
the odd-family (I, III, ...) and even-family (II, IV, ...) of
inviscid/acoustic modes are shown in Fig.~\ref{fig:fig_spectra1}($a$);
the mode-I and mode-II are indicated  in Fig.~\ref{fig:fig_spectra1}($b$).
Note that the `Y'-spectrum for compressible Couette flow 
becomes difficult to resolve numerically 
if $Re$ and $\alpha$ are large~\cite{DEH94,HZ98}.

%-------------------------
%-------------------------

%-------------------------
\begin{figure}[p]
\begin{center}
\begin{tabular}{@{\hspace{-0.2in}}l@{\hspace{-0.65in}}c@{\hspace{-0.65in}}r}
\begin{minipage}[t]{3.0in}
\begin{picture}(3.00,2.5)
\centerline{\psfig{figure=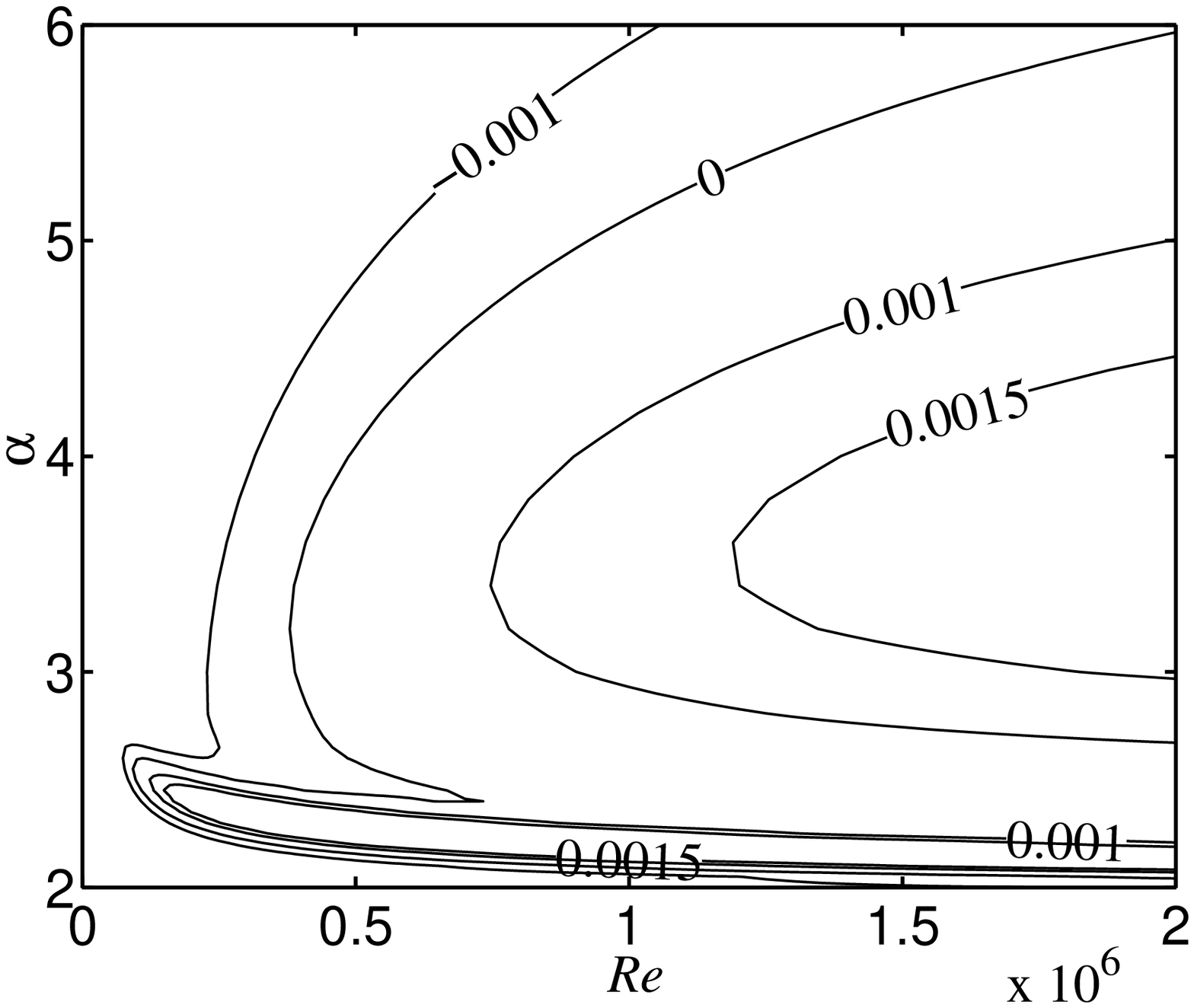,width=3.00in}}
\put(-3.1,2.1){(a)}
\end{picture}
\end{minipage}
\\
\begin{minipage}[t]{3.0in}
\begin{picture}(3.0,2.5)
\centerline{\psfig{figure=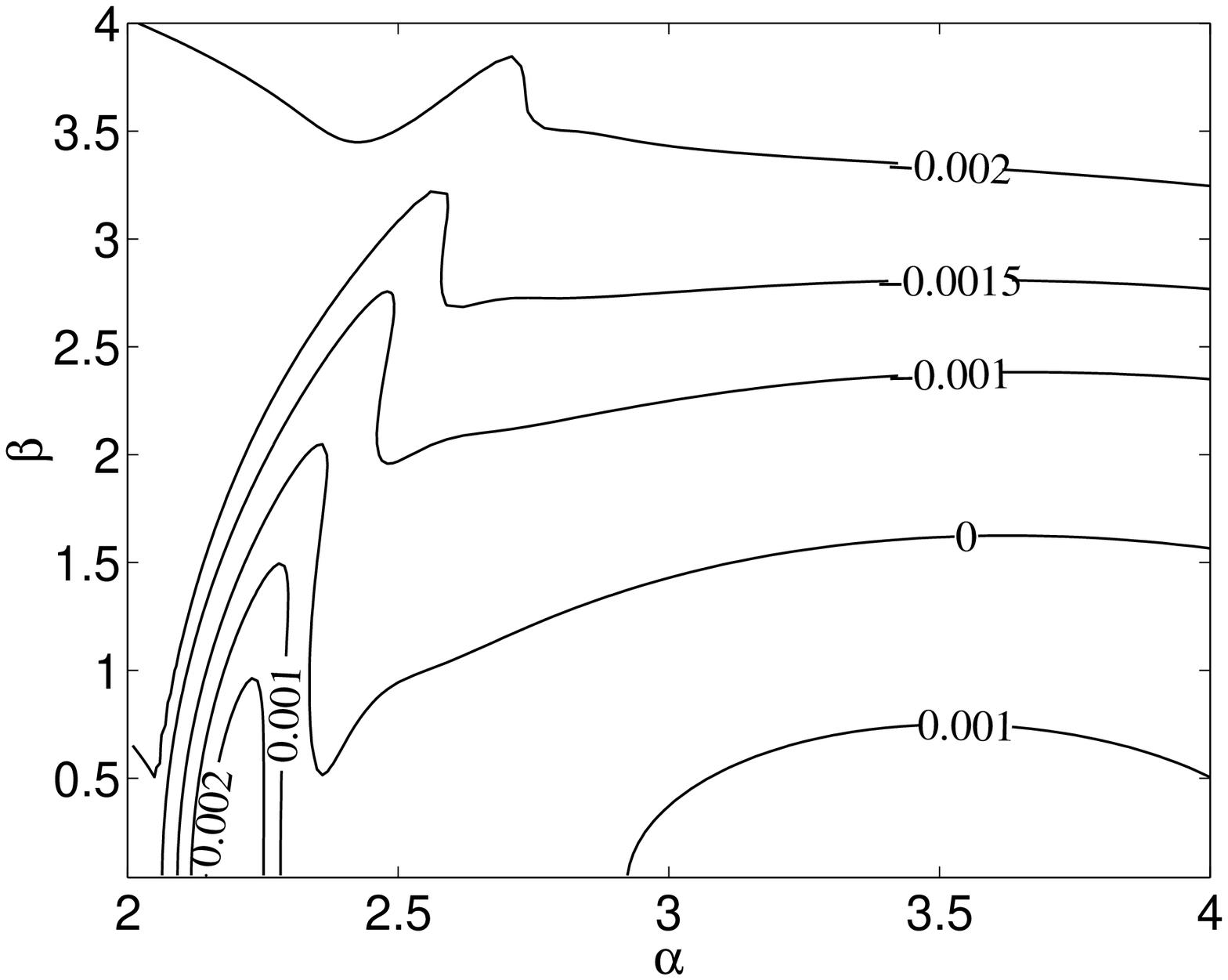,width=3.00in}}
\put(-3.1,2.1){(b)}
\end{picture}
\end{minipage}
\end{tabular}
\end{center}
\caption{
($a$) Contours of the growth rate of the least-decaying
mode, $\omega_{ldi}$, in the ($Re, \alpha$)-plane
with $M=5$ for two-dimensional ($\beta=0$) disturbances.
($b$) Contours of 
$\omega_{ldi}$ in the ($\alpha, \beta$)-plane
with $Re=2\times 10^5$ and $M=5$.
} 
\label{fig:fig_neutral}
\end{figure}
%-------------------------

Now we present a few representative
stability diagrams for compressible plane Couette flow, 
delineating the zones of stability and instability.
Figure \ref{fig:fig_neutral}($a$) shows the contours of the growth rate
of the least decaying mode, $\omega_{ldi}=\max (\omega_i)$,
in the $({\it Re},\alpha)$-plane for 
two-dimensional disturbances ($\beta=0$) at a Mach number of  $M=5$.
(This plot matches  well with Fig. 20$a$ of Hu and Zhong~\cite{HZ98}.)
For 3D disturbances, Fig.~\ref{fig:fig_neutral}($b$) shows the contours of $\omega_{ldi}$
in the ($\alpha, \beta$)-plane for $M = 5$ and ${\it Re}= 2\times 10^5$.
In each plot, the flow is unstable inside the neutral contour,
denoted by `$0$', and stable outside. 
The mode-II (see Fig. 2$b$) is responsible for the observed instabilities; 
for other details on linear stability
results the reader is referred to Hu and Zhong~\cite{HZ98}.

For the transient growth analysis, we will focus on
the control parameter space (such as in Fig.~3)
where the flow is stable according to the modal linear stability analysis,
and investigate the potential of this `stable' flow to give rise 
to transient energy growth.

\section{Transient Energy Growth}

To study the evolution of a general non-eigenstate, we relax the assumption 
of exponential time dependence. The linear perturbation equations can be 
written as~\cite{SH01}
\begin{equation}
  \frac{\partial {\bf \tilde{q}}}{\partial t} = -i {\cal L}{\bf \tilde{q}},
\label{qstateeqn}
\end{equation}
where ${\bf \tilde{q}}(y,t;\alpha,\beta)$ is the inverse Fourier transform of 
$\hat{\bf q}(x,y,z,t)$, as given by~(\ref{modalxzdep}). The state function is 
expanded in the basis of its eigenfunctions as 
\begin {equation}
  {\bf \tilde{q}} = \sum_k \kappa_k(t) {\bf q}'_k,
\label{representation1}
\end{equation}
where the index $k$ runs through a selected portion of the least decaying eigenmodes in the 
complex eigenvalue plane (refer to Fig. 2). 
Using (\ref{representation1}), the equation~(\ref{qstateeqn}) can be diagonalized as
\begin{equation}
   \frac{{\rm d} {\bf \kappa}}{{\rm d}t} = -i{\bf \rm \Omega} {\bf \kappa},
\quad
\mbox{with}
\quad
{\bf \kappa}(t) = e^{-it {\bf \rm \Omega}} {\bf \kappa}(0),
\label{kstateeqn}
\end{equation}
where ${\bf \kappa} = \{\kappa_k\}^{\rm T}$ and ${\bf \rm \Omega} = {\rm diag}\{\omega_k\}$.

To compute the nonmodal energy growth,
we need an expression for the perturbation energy.
Under a suitable definition of the inner product, 
the square of the norm $||{\bf \tilde{q}(t)}||$ can be made to 
represent a measure of this energy density, 
${\cal E}(\alpha, \beta, t)$. 
The energy density is defined as that of Mack~\cite{Mack69,Mack84} 
\begin{equation}
  {\cal E}(\alpha,\beta,t) 
   = \int_{-1}^{1}{\bf \tilde{q}}^{\dagger}(\xi,t)
      {\mathcal M}{\bf \tilde{q}}(\xi,t){\rm d}\xi,
\label{eqn_Macknorm} 
\end{equation}
where the weight matrix,
\begin{equation}
 {\mathcal M} = {\rm diag}\{\rho_0, \rho_0, \rho_0, T_0/\rho_0\gamma M^2, 
  \rho_0/\gamma(\gamma-1)T_0M^2 \},
\label{eqn_WMatrix1}
\end{equation}
is diagonal.
In this definition of ${\cal E}(\alpha,\beta,t)$, the spatial average of the 
rate of pressure-related work (i.e. the compression work) is zero. 
As elaborated by Mack~\cite{Mack84},
the contribution of this compression work 
to the total perturbation energy should vanish due to the conservative
nature of the compression work.
Note that the Mach energy norm (\ref{eqn_Macknorm}) has also been used
in the `spatial' transient growth analyses of compressible flows~\cite{TR03}.

Let $G(t, \alpha,\beta; Re, M)$ be the amplification of the initial energy 
density maximized over all possible initial conditions, i.e., 
\begin{eqnarray}
   G(t, \alpha,\beta; Re, M) 
    \equiv G(t) &=& 
    \max_{\bf \tilde{q}(0)} \frac{{\cal E}(\alpha,\beta,t)}{{\cal E}(\alpha,\beta,0)} .
\label{GeqnL2norm}
\end{eqnarray}
For the numerical evaluation of $G(t)$, we refer the readers to Refs.~\cite{RH93,SH01,HSH96}.

%-------------------------
\begin{figure}[p]
\begin{center}
\begin{tabular}{lr}
\begin{minipage}[t]{3.0in}
\begin{picture}(3.0,2.5)
\centerline{\psfig{figure=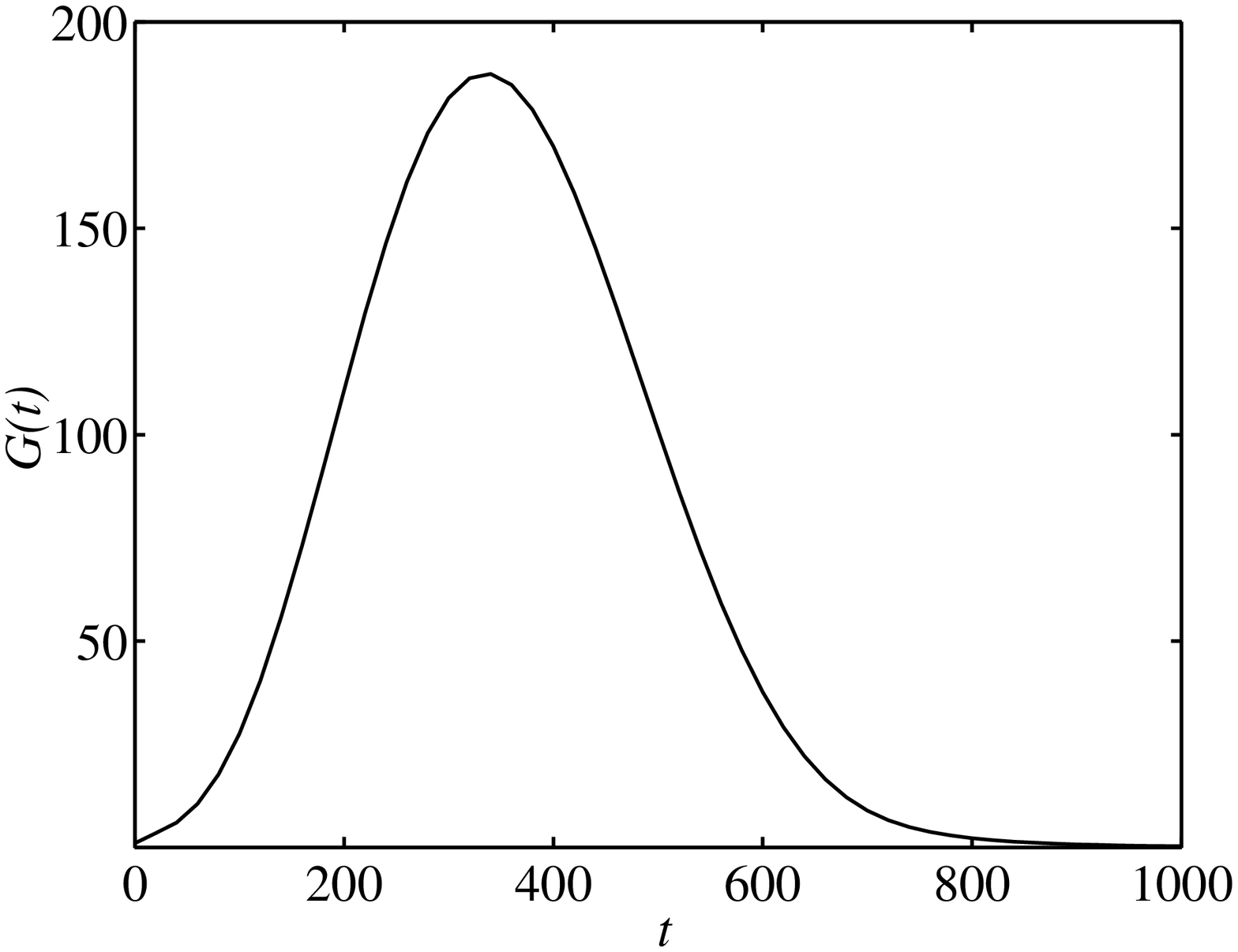,width=3.0in}}
\put(-3.1,2.1){(a)}
\end{picture}
\end{minipage}
\\ 
\begin{minipage}[t]{3.0in}
\begin{picture}(3.0,2.5)
\centerline{\psfig{figure=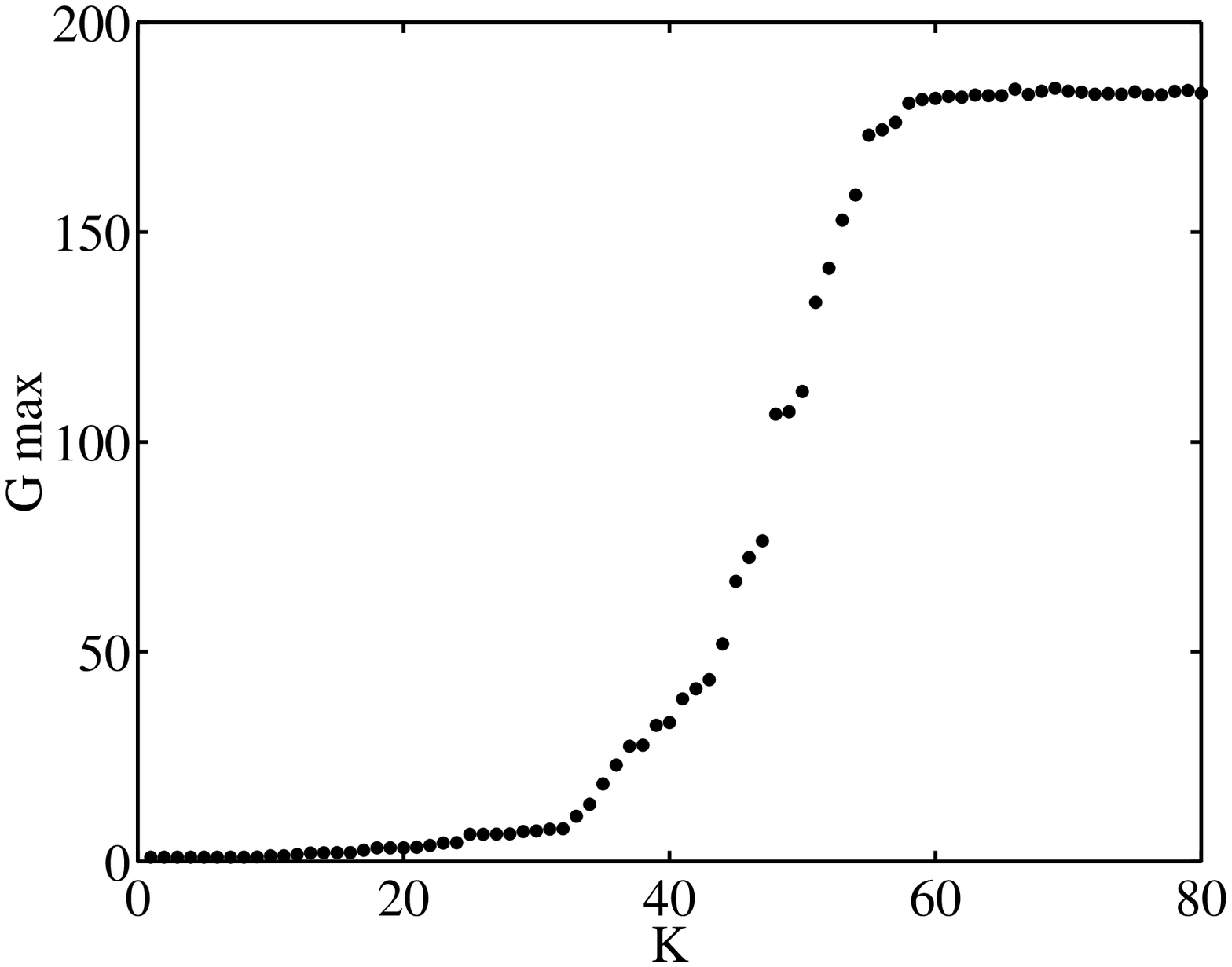,width=3.0in}}
\put(-3.1,2.1){(b)}
\end{picture}
\end{minipage}
\end{tabular}
\end{center}
\caption{(a) Variation of the energy amplification factor, $G(t)$, with time.
(b) Dependence of the maximum energy growth, $G_{max}$, in panel $a$ with
the number of modes, $K$, considered ($N=100$).
Parameter values are ${\it Re} = 2\times 10^5$, $M = 2$ and 
$\alpha = 0.1=\beta$.
}
\label{GtM2R2e5al01bt01}
\end{figure}
%-------------------------

A typical variation of the energy amplification factor, $G(t, \alpha,\beta)$,
with time is shown in  Fig.~\ref{GtM2R2e5al01bt01}($a$),
for the parameter values of $\alpha = 0.1=\beta$, ${\it Re} = 2\times 10^5$ and $M = 2$.
There is a 200-fold increase of the initial energy density at $t\approx 300$.
Since the flow is stable for this parameter combination, 
$G(t)$ decays in the asymptotic limit ($t\to \infty$). 
To compute $G(t)$ we do not need to consider all modes
since we have found that
the modes with large phase speeds (see Fig. 2) do not contribute to the energy growth.  
This observation is similar to the earlier findings 
for compressible boundary layers~\cite{HSH96}.
Therefore, to reduce the computational effort, 
the index $k$ in (\ref{representation1}) is chosen corresponding to 
the modes whose phase speeds are within the range $-1 < \omega_r/\alpha < 2$
(i.e., comparable to the extremes of the mean flow velocity which varies
between $0$ and $1$), 
and the decay rate is less than $0.5$ (i.e., $\omega_i > -0.5$). 
With this choice of modes,
the maximum possible energy amplification  over time,
\begin{equation}
   G_{\rm max}(\alpha,\beta; Re, M) = \max_{t\ge 0} G(t,\alpha,\beta; Re, M)
\quad \mbox{at} \quad t=t_{max},
\end{equation}
saturates to some constant value with the number of modes
retained $K$, as shown in Fig.~\ref{GtM2R2e5al01bt01}($b$). 
It is observed that out of a total of $500$ modes, only $70$ 
modes are sufficient to compute $G_{\rm max}$ accurately,
and the contribution of the remaining modes to $G_{max}$ is negligibly small.

%-------------------------
\begin{figure}[p]
\begin{center}
\begin{tabular}{c}
\begin{minipage}[t]{3.0in}
\begin{picture}(3.0,2.5)
\centerline{\psfig{figure=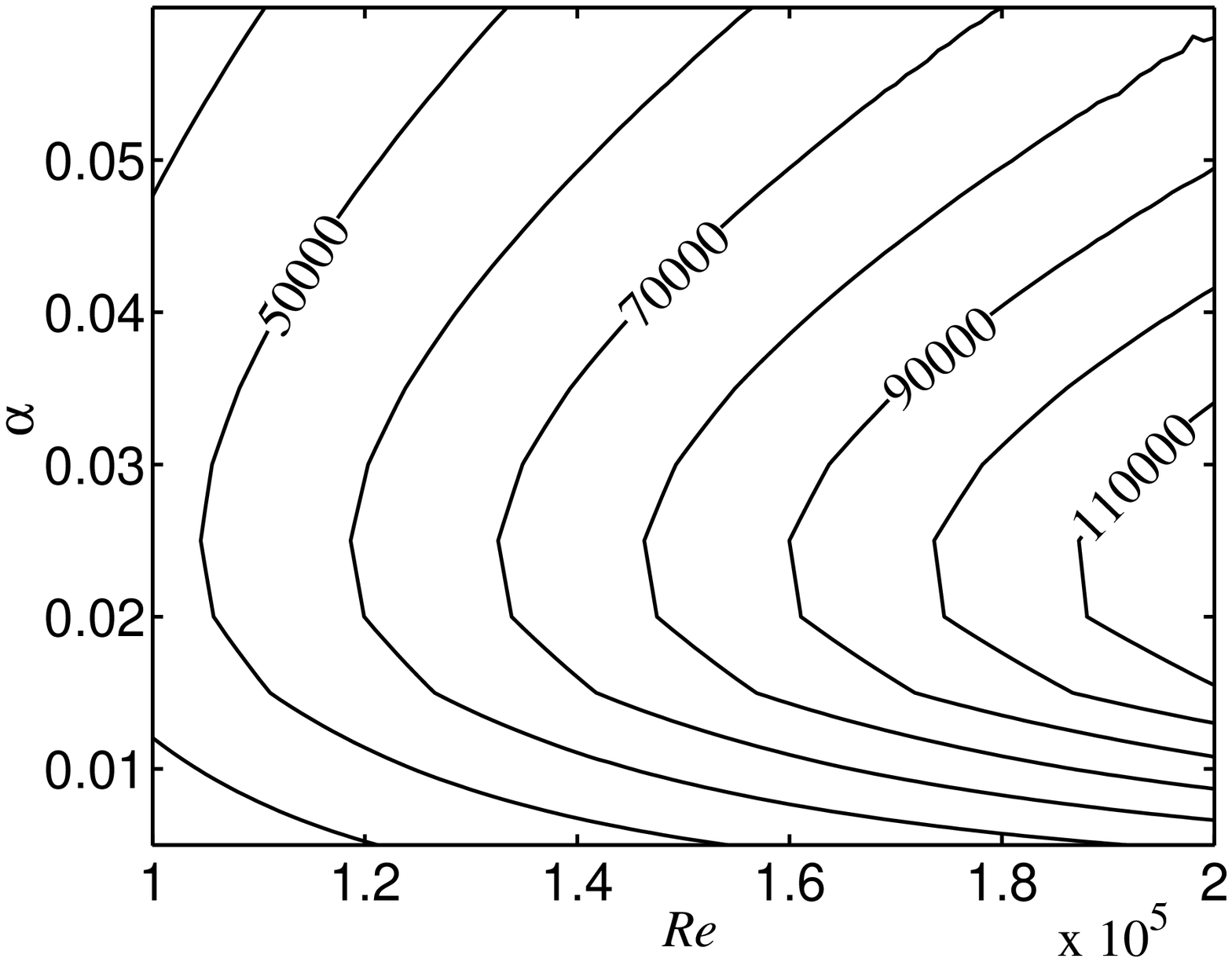,width=3.0in}}
\end{picture}
\end{minipage}
\end{tabular}
\end{center}
\caption{Contours of maximum transient growth, $G_{max}$, in the (${\it Re}, \alpha$)-plane 
for $M = 3$ and $\beta = 3$.
} 
\label{GRxalM3bt3_MxalphaR2e5bt3}
\end{figure}
%-------------------------

For $M=3$ and $\beta=3$,
the contours of $G_{max}$ in the (${\it Re}, \alpha$)-plane are shown 
in Fig.~\ref{GRxalM3bt3_MxalphaR2e5bt3}($a$).
For other values of $M$ and $\beta$, the $G_{max}$-contours look similar. 
It is observed that $G_{max}$ varies non-monotonically
with streamwise wavenumber $\alpha$ for  fixed $Re$,
and is maximum for non-zero values of $\alpha$;
$G_{max}$ increases with Reynolds number for any value of  $\alpha$. 
The latter observation mirrors similar findings on transient growth
in incompressible fluids~\cite{SH01}.

%-------------------------
\begin{figure}[p]
\begin{center}
\begin{tabular}{c}
\begin{minipage}[t]{3.0in}
\begin{picture}(3.0,2.5)
\centerline{\psfig{figure=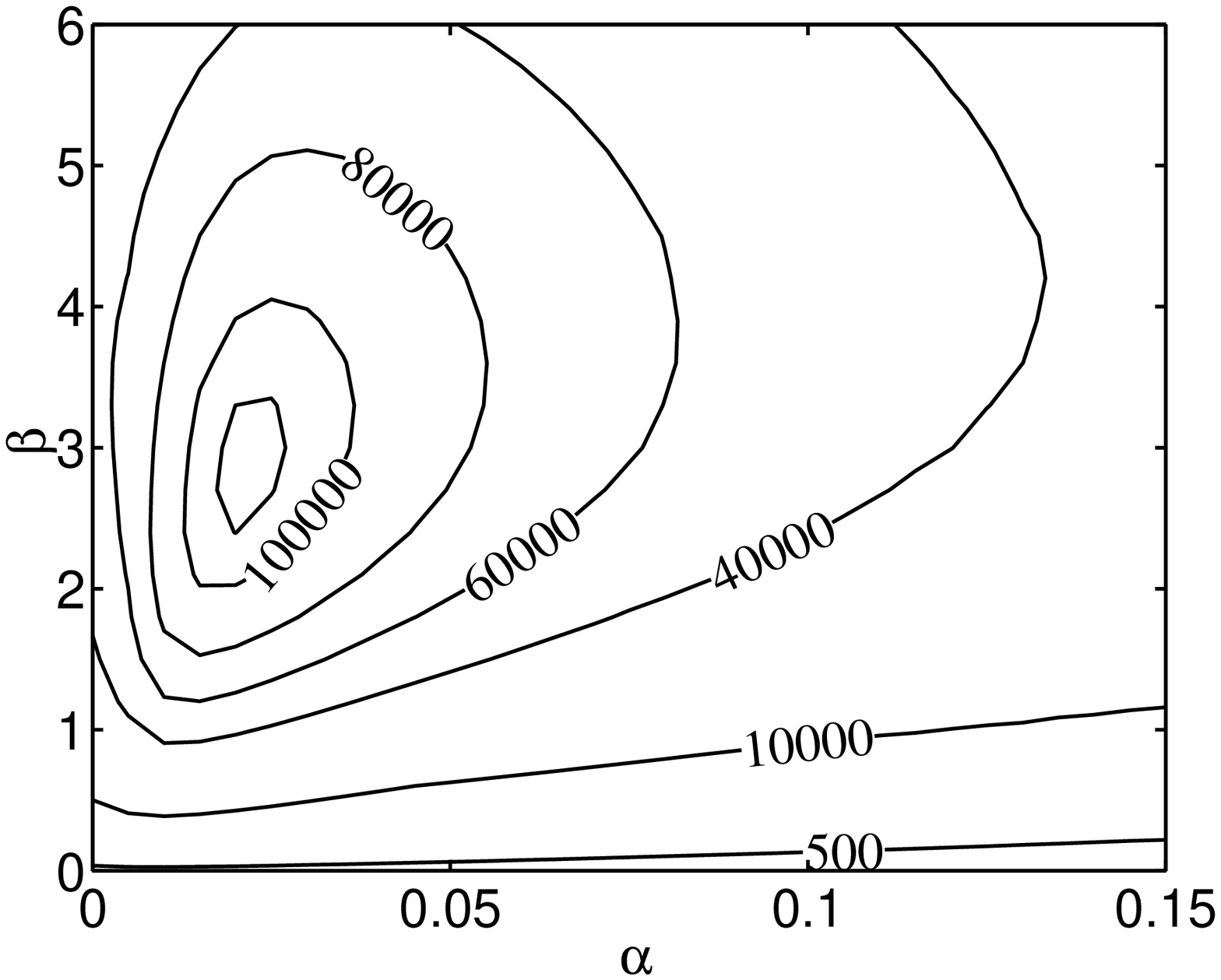,width=3.0in}}
\put(-3.1,2.1){(a)}
\end{picture}
\end{minipage}
%\\
\begin{minipage}[t]{3.0in}
\begin{picture}(3.0,2.5)
\centerline{\psfig{figure=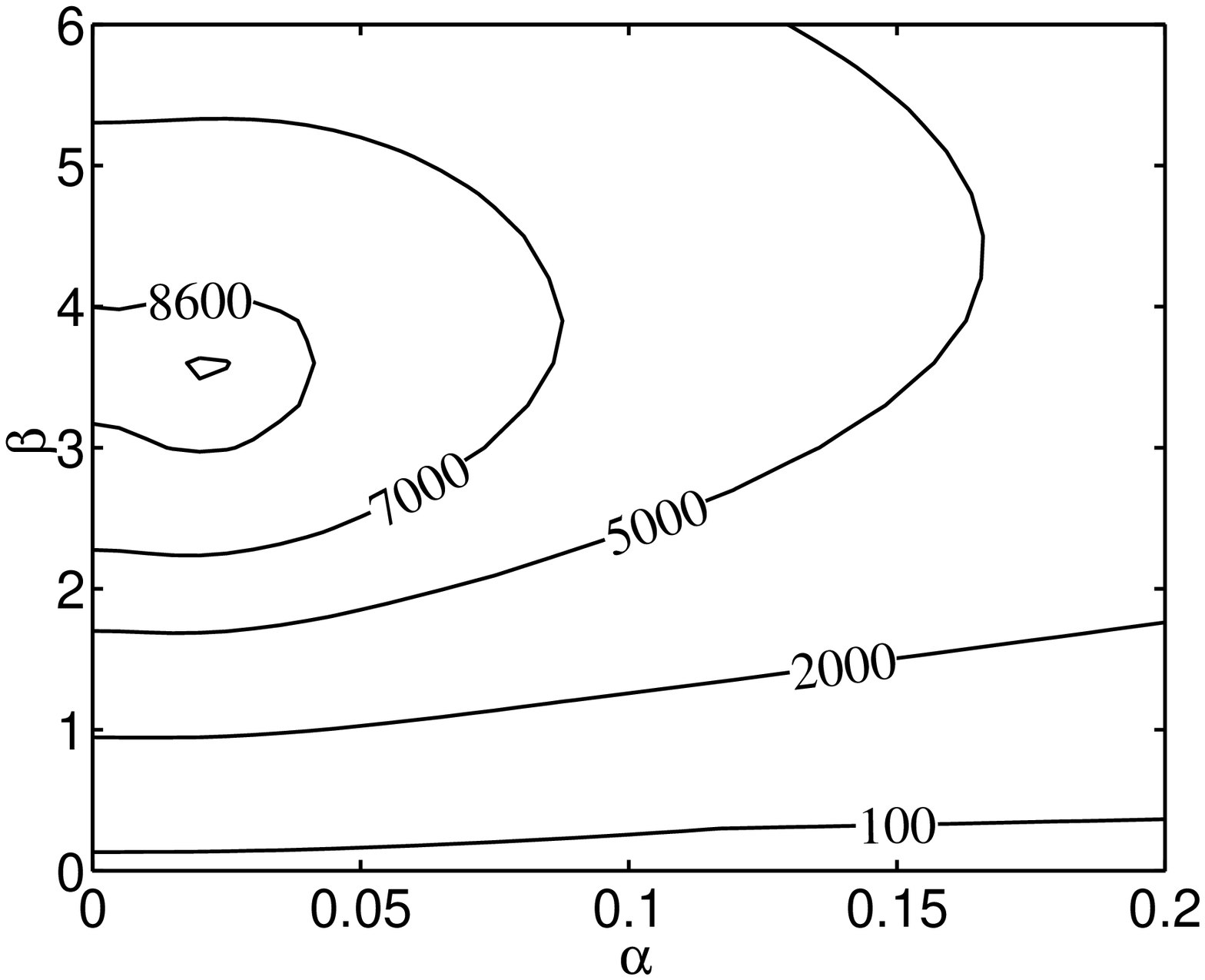,width=3.0in}}
\put(-3.1,2.1){(b)}
\end{picture}
\end{minipage}
\end{tabular}
\end{center}
\caption{Contours of $G_{max}$ in the ($\alpha, \beta$)-plane
for ${\it Re} = 10^5$ at various Mach numbers: 
(a) $M = 2$; (b) $M = 5$.
}
\label{GalxbtR1e5}
\end{figure}
%-------------------------

Moving onto the effect of Mach number on the transient energy growth,
we show, in Fig.~\ref{GalxbtR1e5}($a$) and \ref{GalxbtR1e5}($b$),
the contours of $G_{\max}$ in the ($\alpha, \beta$)-plane for 
$M = 2$ and $5$, respectively;
the Reynolds number is set to $Re=10^5$.
It is observed that the magnitude of $G_{\max}$ decreases
sharply as we increase the Mach number from $2$ to $5$
for any combination of $\alpha$ and $\beta$. 
The supremum  of $G_{max}$ over all possible combinations
of wavenubers $(\alpha,\beta)$ 
is called the optimal energy growth, $G_{\rm opt}$:
\begin{equation}
   G_{\rm opt}(Re, M) = \sup_{\alpha,\beta} G_{max}(\alpha,\beta; Re, M),
\end{equation}
which corresponds to $(t_{\rm opt},\alpha_{\rm opt},\beta_{\rm opt})$. 
In Fig.~\ref{GalxbtR1e5}($a,b$), 
this global maximum is seen to occur at $\alpha\sim 0$ and $\beta\sim 3$.
Also, $G_{\rm opt}$  decreases 
by an order of magnitude with increasing Mach number
as seen in Figs. ~\ref{GalxbtR1e5}($a$) and ~\ref{GalxbtR1e5}($b$). 
(An explanation for this behaviour is provided in the next section, along with 
a detailed energy analysis in terms of different  constituent energies.) 
It is worth pointing out that our result on the decrease of
$G_{max}$ with $M$ is in contrast
to that for boundary layers~\cite{HSH96}.

%-------------------------
\begin{figure}[p]
\begin{center}
\begin{tabular}{c}
\begin{minipage}[t]{3.0in}
\begin{picture}(3.0,2.5)
\centerline{\psfig{figure=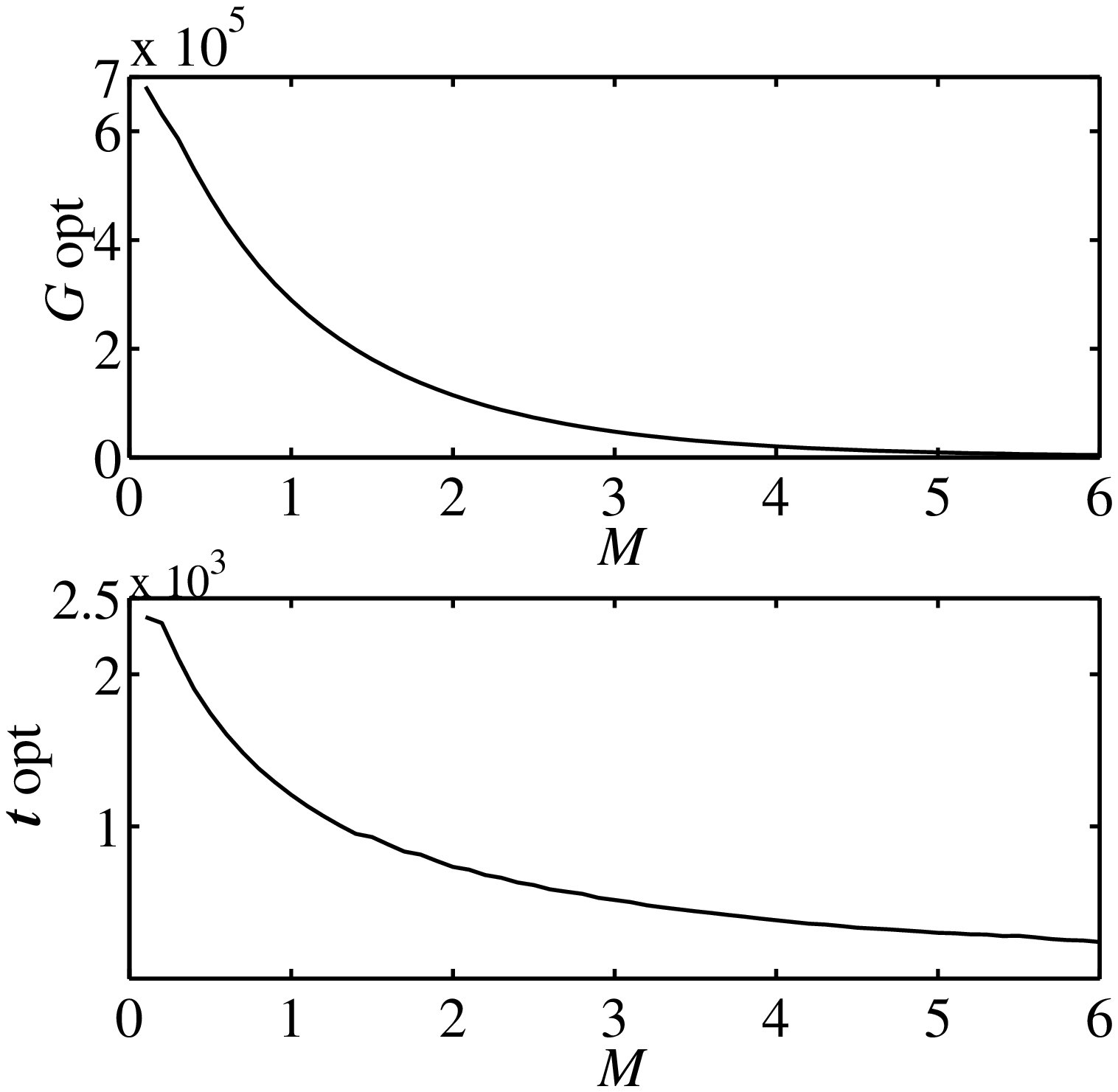,width=3.0in}}
\put(-.5,2.2){(a)}
\put(-.5,0.9){(b)}
\end{picture}
\end{minipage}
%\\
\begin{minipage}[t]{3.0in}
\begin{picture}(3.0,2.5)
\centerline{\psfig{figure=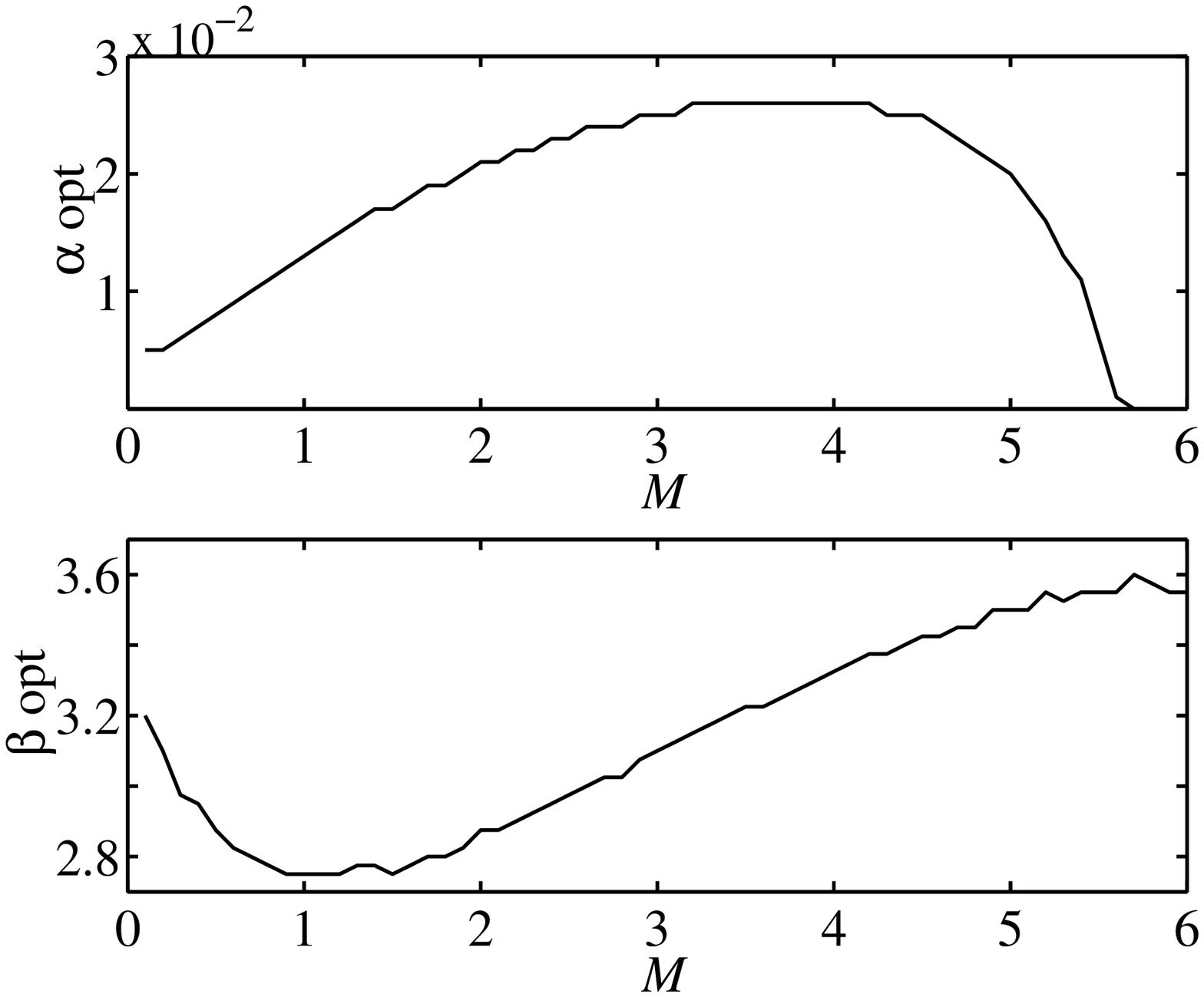,width=3.0in}}
\put(-.5,2.1){(c)}
\put(-.5,0.7){(d)}
\end{picture}
\end{minipage}
\end{tabular}
\end{center}
\caption{Variations of ($a$) the optimal energy growth, $G_{\rm opt}$, 
($b$) optimal time, $t_{\rm opt}$, ($c$) optimal streamwise wavenumber, $\alpha_{\rm opt}$, and 
($d$) optimal spanwise wavenumber, $\beta_{\rm opt}$,
 with the Mach number $M$ for ${\it Re} = 10^5$. 
} 
\label{GtalbtoptxMRe1e5}
\end{figure}
%-------------------------

The variations of the optimal growth $G_{\rm opt}$, 
the corresponding optimal time $t_{\rm opt}$, 
the optimal streamwise, $\alpha_{\rm opt}$, and 
transverse, $\beta_{\rm opt}$, wavenumbers with 
the Mach number  are shown in Fig.~\ref{GtalbtoptxMRe1e5}($a$-$d$);
as in Fig. 6, $Re= 10^5$ for these plots. It is observed 
that while both $G_{\rm opt}$ and $t_{\rm opt}$ decrease monotonically
with increasing $M$,
both the optimal wavenumbers ($\alpha_{\rm opt}$ and $\beta_{\rm opt}$)
vary nonmonotonically in the same limit.
(Note that the wiggles in panels $c$ and $d$ are due to the
searching of $G_{\rm opt}$ in the ($\alpha, \beta$)-plane
with small but finite grid-spacing-- these computations are very time consuming.)
Figure \ref{GtalbtoptxMRe1e5}($c$) shows the interesting result that the 
optimal streamwise wavenumber $\alpha_{\rm opt}$ can become zero at large $M$.
This is in contrast to the incompressible 
Couette flow for which  $\alpha_{\rm opt}$ 
stays always slightly away from the null value~\cite{SH01}.

\subsection{Structure of optimal disturbances}

The  initial  disturbance pattern 
for  given values of $\alpha$ and $\beta$ that reaches the maximum possible transient 
growth $G_{\max}$ at $t_{\max}$ along with 
the corresponding disturbance pattern at time $t_{\max}$ can be determined through 
the singular value decomposition~\cite{SH01}.
In the following, such disturbances  at $t=0$ and $t_{\max}$
are referred to as the {\it optimal} disturbances.

%-------------------------
\begin{figure}[p]
\centerline{\psfig{figure=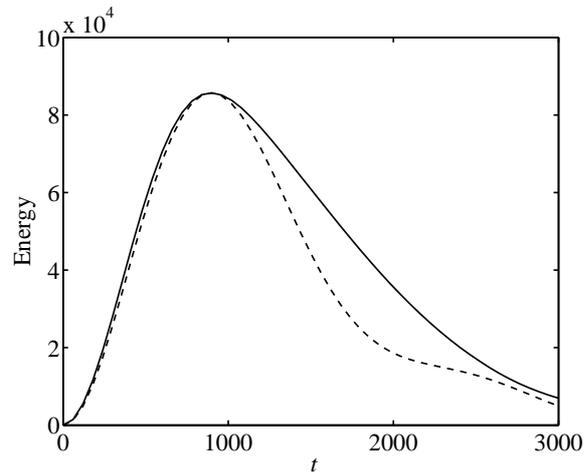,width=3.0in}}
\caption{
Variations of $G(t)$ (solid line) and ${\mathcal E}(t)/{\mathcal E}(0)$ 
(dash line) with time 
for $\alpha = 0$, $\beta = 3$, $M = 2$ and ${\it Re} = 2 \times 10^5$. 
}
\label{GtEtal0bt3R2e5M2}
\end{figure}
%-------------------------

Before presenting  the structure of optimal velocity patterns,
we show the variation of ${\mathcal E}(t)/{\mathcal E}(0)$ with time,
denoted by the dash line, in Fig. ~\ref{GtEtal0bt3R2e5M2} for the particular
initial perturbation that reaches maximum energy, 
$G_{\max}$, at a time $t_{\max}$. 
The parameter values are set to  $M=2$, $Re=2\times10^5$,
$\alpha=0$ and $\beta=3$.  For comparison, we have
also superimposed the variation of $G(t)$ with time, denoted by the solid line.
As expected, the envelope of ${\mathcal E}(t)/{\mathcal E}(0)$ is
below that of $G(t)$ since the former corresponds to a particular
initial condition but the latter to all possible combinations of
initial conditions that yields maximum amplification at any time.

%-------------------------
\begin{figure}[p]
\begin{center}
\begin{tabular}{c}
\begin{minipage}[t]{3.0in}
\begin{picture}(3.0,2.5)
\centerline{\psfig{figure=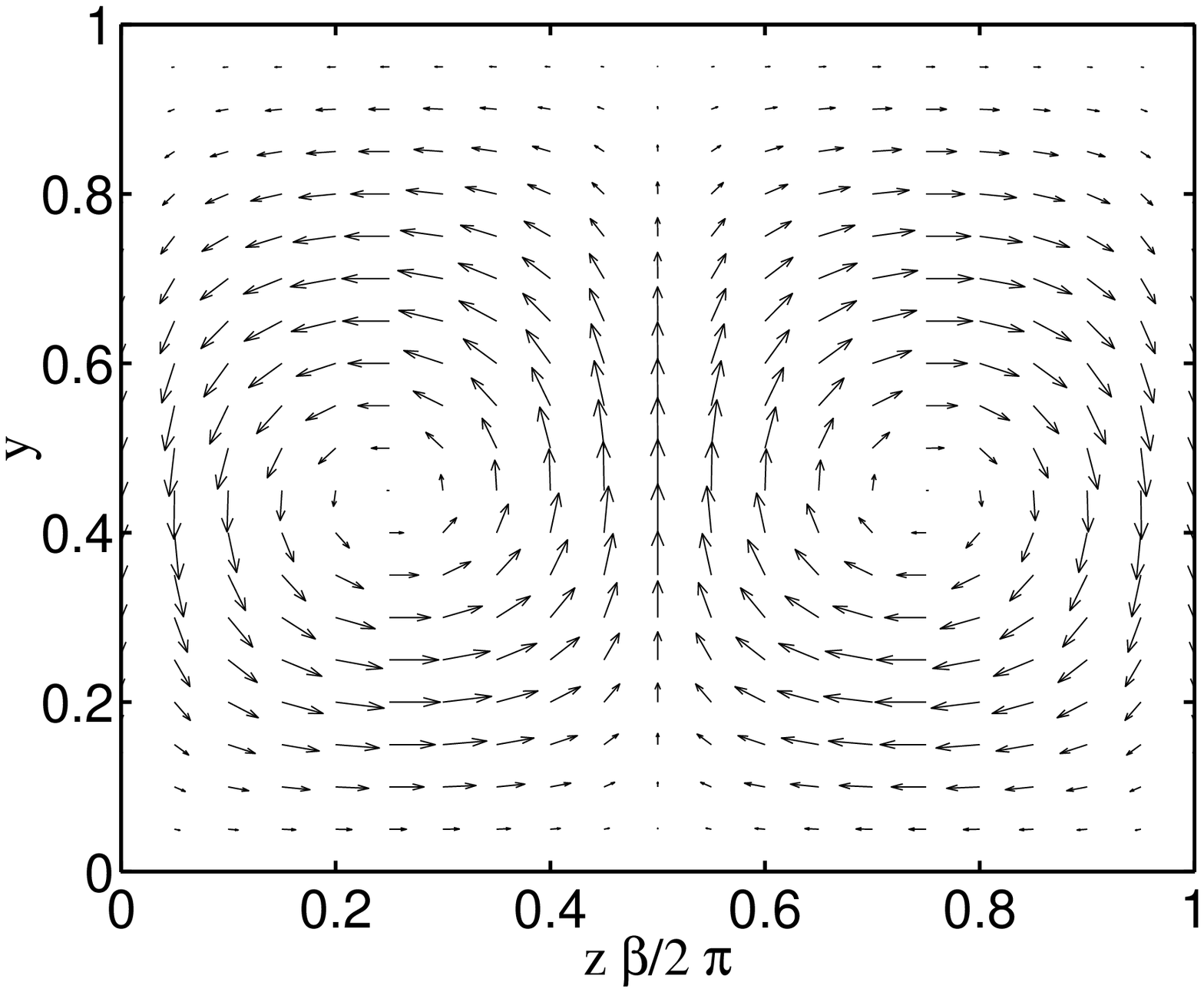,width=3.0in}}
\put(-3.1,2.1){(a)}
\end{picture}
\end{minipage}
\\
\begin{minipage}[t]{3.0in}
\begin{picture}(3.0,2.5)
\centerline{\psfig{figure=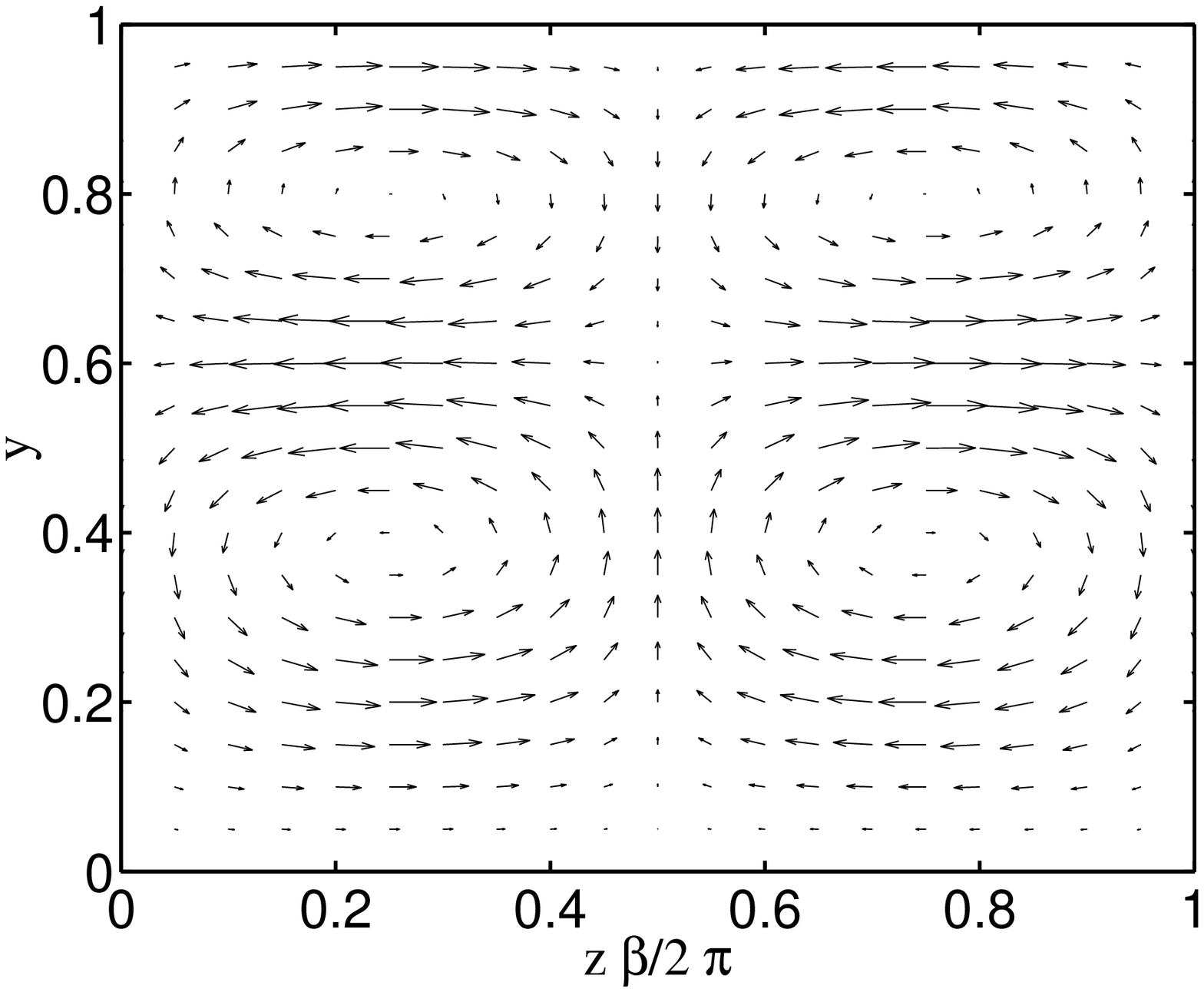,width=3.0in}}
\put(-3.1,2.1){(b)}
\end{picture}
\end{minipage}
\\
\begin{minipage}[t]{3.0in}
\begin{picture}(3.0,2.5)
\centerline{\psfig{figure=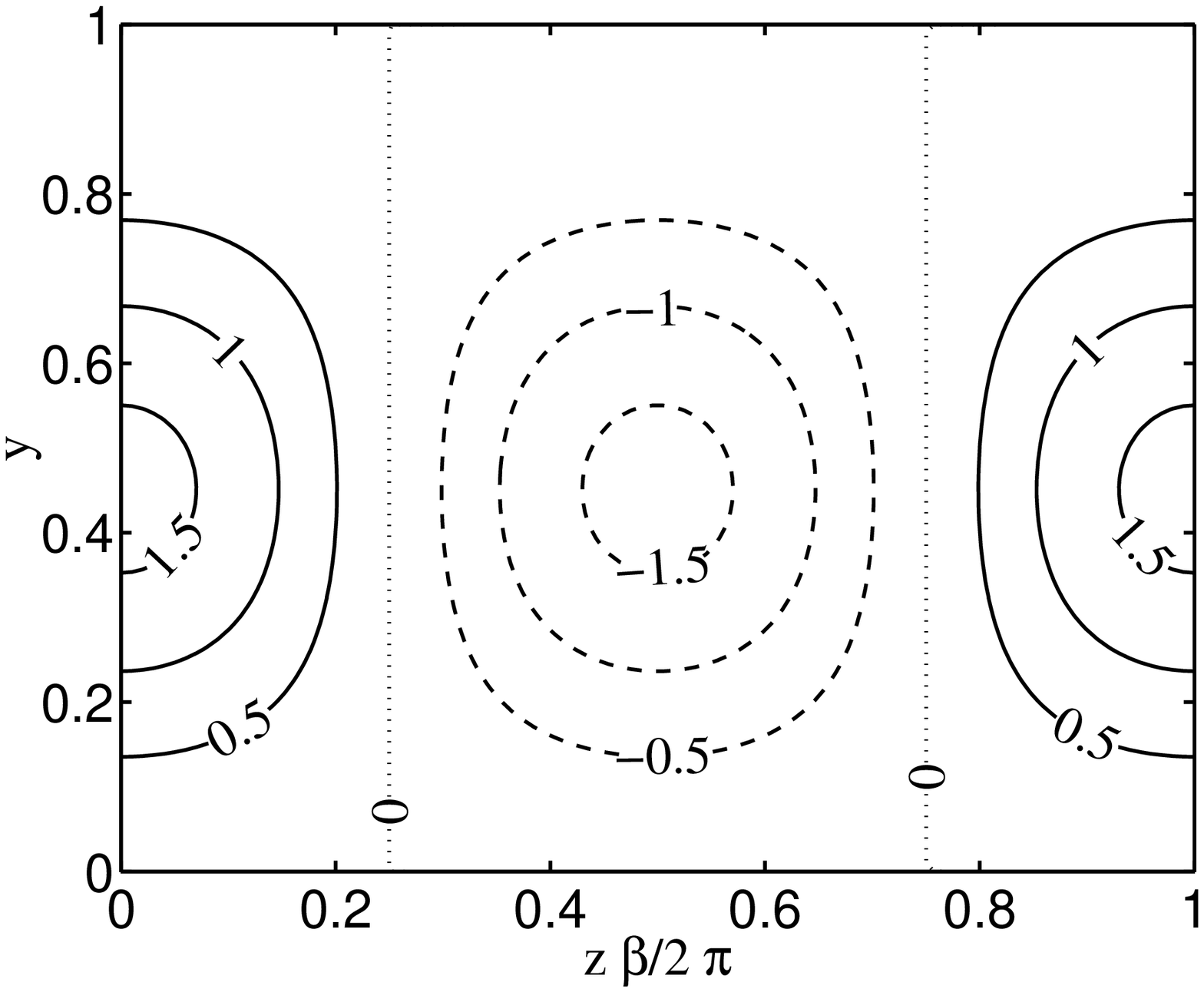,width=3.0in}}
\put(-3.1,2.1){(c)}
\end{picture}
\end{minipage}
\end{tabular}
\end{center}
\caption{Optimal patterns of perturbation velocity at ($a$)  $t=0$ and ($b$) $t=t_{max}$ 
in the ($z, y$)-plane for $\alpha = 0$, 
$\beta = 3$, $M = 2$ and ${\it Re} = 2\times 10^5$.
($c$) Contours of optimal streamwise perturbation velocity 
at $t=t_{max}$ for the same parameter set.
} 
\label{wvM2R1e5al0bt3}
\end{figure}
%-------------------------

For the maximum energy growth as in Fig.~\ref{GtEtal0bt3R2e5M2},
the optimal velocity patterns at $t=0$ and $t_{max}$ in the $(y, z)$-plane
are shown in Fig.~\ref{wvM2R1e5al0bt3}($a$) and 
Fig.~\ref{wvM2R1e5al0bt3}($b$), respectively.
(Note that these patterns are invariant along the
streamwise direction since $\alpha=0$.)
Both patterns resemble streamwise vortices which
are typical of all shear flows.
It is noteworthy that  the optimal pattern in Fig.~\ref{wvM2R1e5al0bt3}($b$)
has two counter-rotating streamwise vortices along the wall-normal direction.
The inviscid rapid growth of these optimal patterns  for 
null or very low values of the streamwise wavenumber ($\alpha$) is due to
the {\it lift-up mechanism}~\cite{Land80} during the 
motion of the fluid particles in the wall-normal 
direction. Another typical feature of shear flows is 
the spanwise alternating streamwise velocities as shown in  
Fig.~\ref{wvM2R1e5al0bt3}($c$). 

%-------------------------
\begin{figure}[p]
\begin{center}
\begin{tabular}{c}
\begin{minipage}[t]{3.0in}
\begin{picture}(3.0,2.5)
\centerline{\psfig{figure=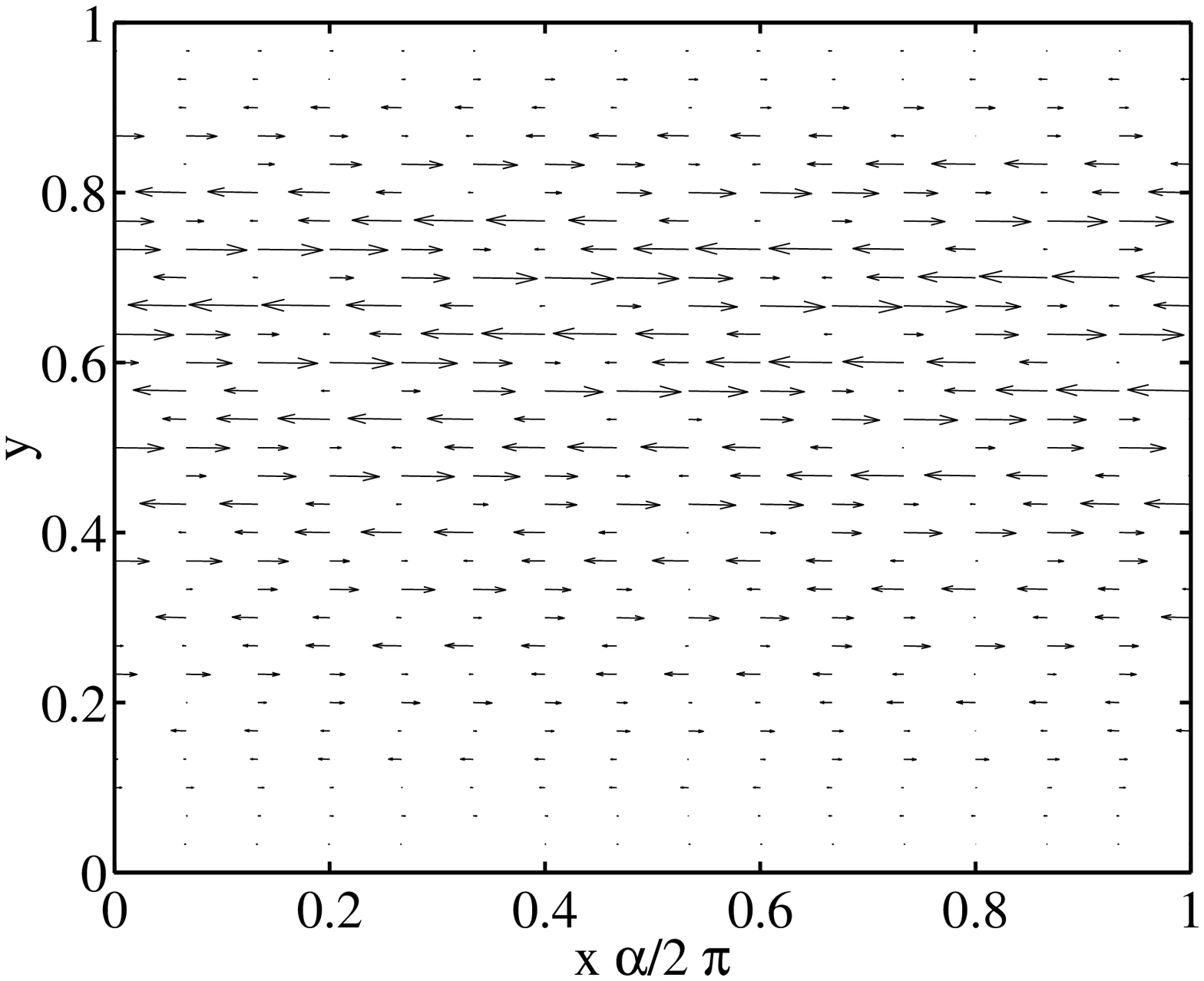,width=3.0in}}
\put(-3.1,2.1){(a)}
\end{picture}
\end{minipage}
\\
\begin{minipage}[t]{3.0in}
\begin{picture}(3.0,2.5)
\centerline{\psfig{figure=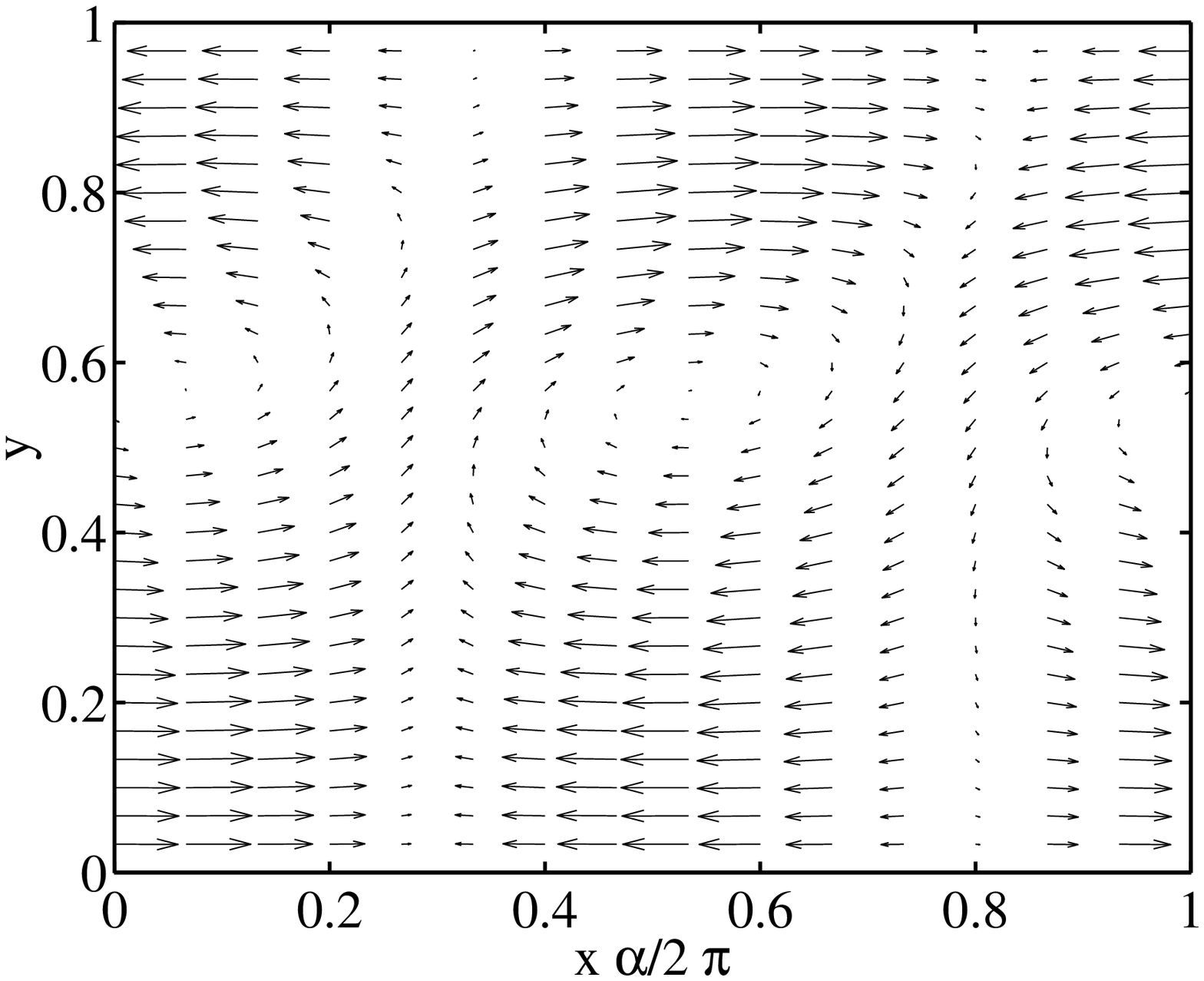,width=3.0in}}
\put(-3.1,2.1){(b)}
\end{picture}
\end{minipage}
\end{tabular}
\end{center}
\caption{Optimal velocity patterns at ($a$)  $t=0$ and ($b$) $t=t_{max}$ 
in the ($x, y$)-plane for two-dimensional disturbances ($\beta = 0$)
with $\alpha = 1$, $M = 2$ and ${\it Re} = 2\times 10^5$.
} 
\label{uvM2R1e5al1bt0}
\end{figure}
%-------------------------

Figures~\ref{uvM2R1e5al1bt0}($a$) and ~\ref{uvM2R1e5al1bt0}($b$) show 
representative two-dimensional ($\beta=0$) optimal velocity patterns in the $(x,y)$ plane.
The parameter values are set to $\alpha=1$, $M=2$ and $Re=2\times 10^5$.
It is observed that the pattern at $t=0$ has structures that 
locally oppose the mean shear.
This pattern subsequently evolves into two `cross-flow' vortices at $t=t_{max}$.
This final configuration is the outcome of the well-known Orr-mechanism~\cite{Orr07}
that leads to transient energy growth due to the tilting of
the initial perturbations into the direction of the mean shear as time evolves.

%-------------------------
\begin{figure}[p]
\begin{center}
\begin{tabular}{c}
\begin{minipage}[t]{3.0in}
\begin{picture}(3.0,2.5)
\centerline{\psfig{figure=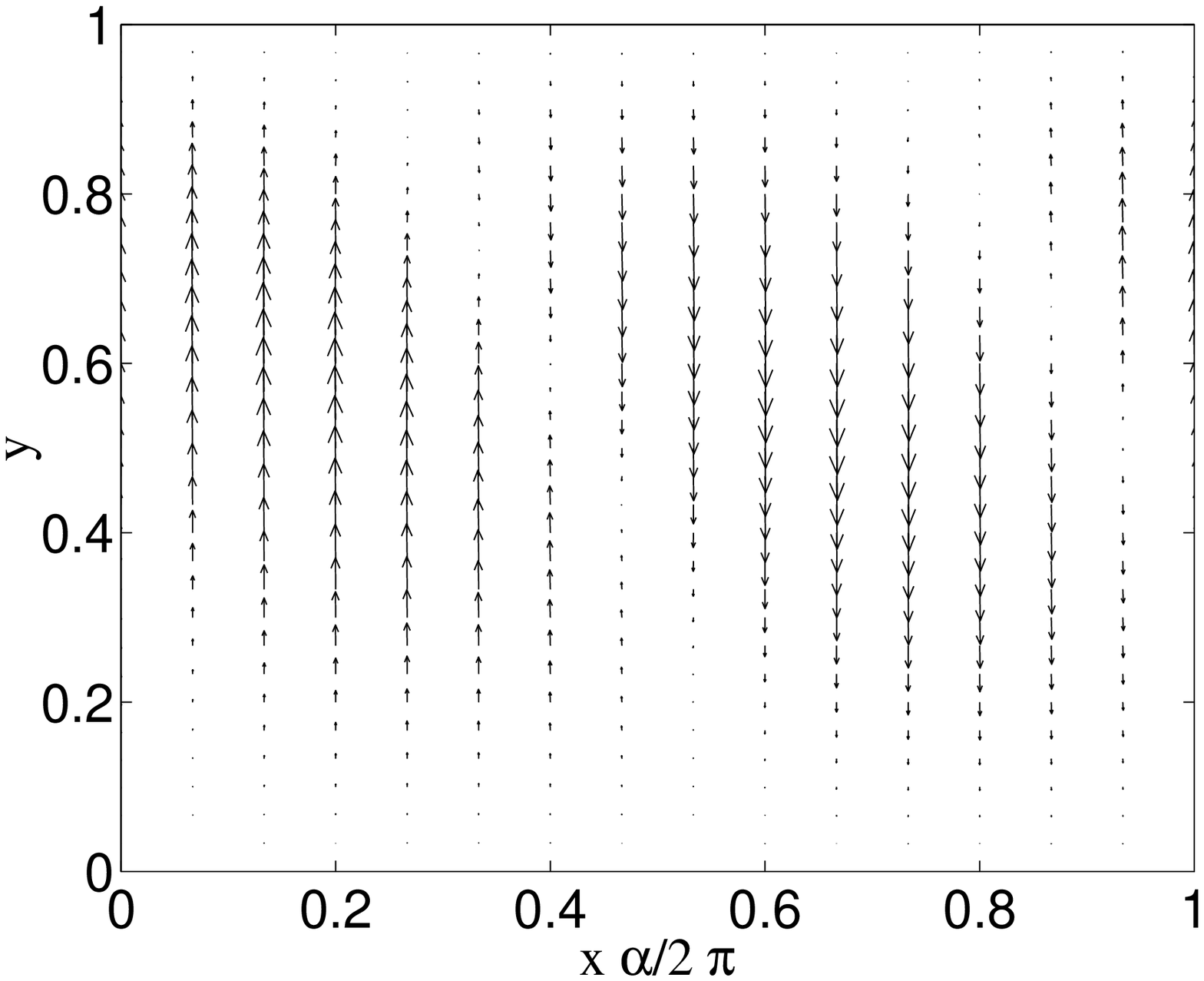,width=3.0in}}
\put(-3.0,2.1){(a)}
\end{picture}
\end{minipage}
\begin{minipage}[t]{3.0in}
\begin{picture}(3.0,2.5)
\centerline{\psfig{figure=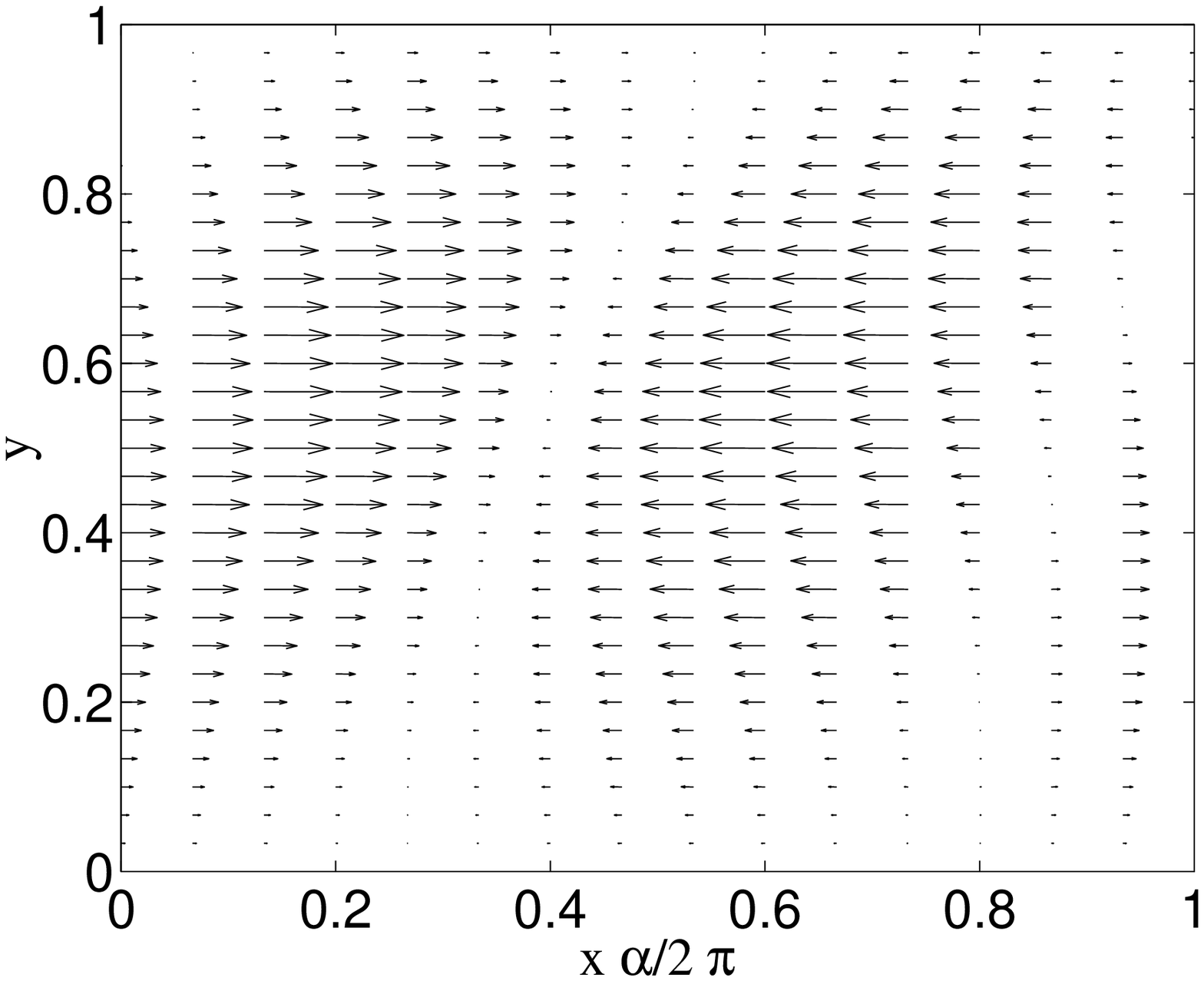,width=3.0in}}
\put(-3.0,2.1){(b)}
\end{picture}
\end{minipage}
\\
\begin{minipage}[t]{3.0in}
\begin{picture}(3.0,2.5)
\centerline{\psfig{figure=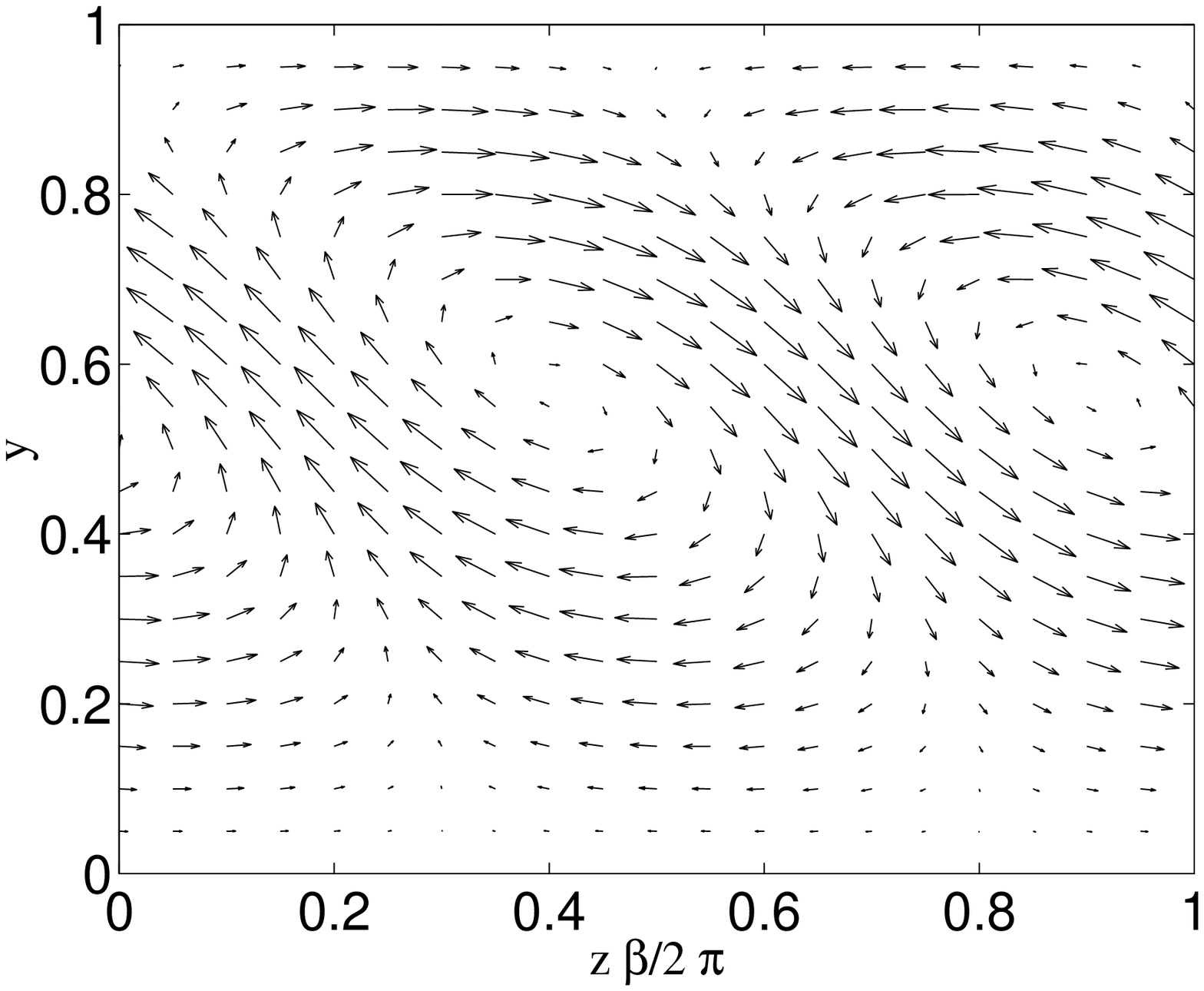,width=3.0in}}
\put(-3.0,2.1){(c)}
\end{picture}
\end{minipage}
\begin{minipage}[t]{3.0in}
\begin{picture}(3.0,2.5)
\centerline{\psfig{figure=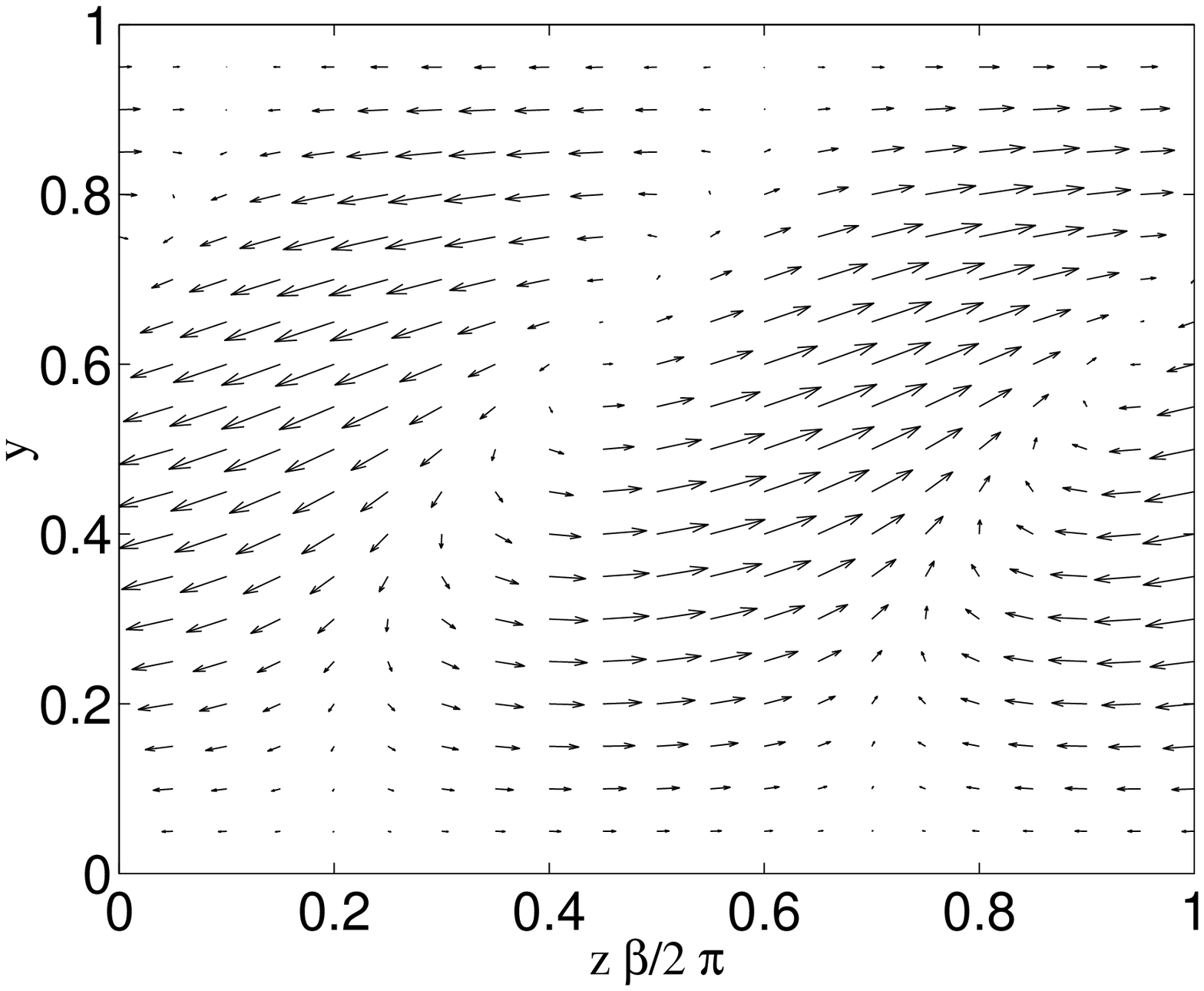,width=3.0in}}
\put(-3.0,2.1){(d)}
\end{picture}
\end{minipage}
\end{tabular}
\end{center}
\caption{Optimal velocity patterns at ($a$)  $t=0$ and ($b$) $t=t_{max}$ 
in the ($x, y$)-plane at $z=0$;
optimal velocity patterns at ($c$)  $t=0$ and ($d$) $t=t_{max}$  
in the ($z, y$)-plane at $x=0$.
Parameter values correspond to the optimal energy growth $G_{opt}$ in
Fig. 7:
$M=5$, $\alpha = 0.02$, $\beta=3.5$ and ${\it Re} = 10^5$.
} 
\label{fig:fig_opt_M5R1e5}
\end{figure}
%------------------------------------

Returning to  the variation of the optimal wavenumber, $\alpha_{opt}$,
with the Mach number in Fig.~\ref{GtalbtoptxMRe1e5}($c$), we note that 
$\alpha_{opt}$ is very small but finite at low-to-moderate values of $M$
and zero at some large value of $M$.
This implies that the {\it pure streamwise vortices} are the optimal patterns 
at high Mach numbers, but their modulated cousins 
are the optimal patterns for low-to-moderate values of the Mach number.
Our results on optimal patterns at high Mach numbers ($M>5.5$)
should be contrasted with that for incompressible Couette flow
for which the {\it oblique modes} constitute optimal patterns.
Typical optimal patterns of such modulated streamwise vortices are shown in
the $(x,y)$- and $(z,y)$-plane 
in Fig.~\ref{fig:fig_opt_M5R1e5} for the optimal energy growth ($G_{\rm opt}$)
at $M=5$, with other parameters as in Fig. 7.
We conclude that the structural features of the optimal disturbance patterns
in compressible Couette flow look similar to those of incompressible shear flows.

\subsection{Scalings of $G_{max}$ and $t_{max}$}

\begin{figure}[p]
\begin{center}
\begin{tabular}{c}
\begin{minipage}[t]{3.0in}
\begin{picture}(3.0,2.5)
\centerline{\psfig{figure=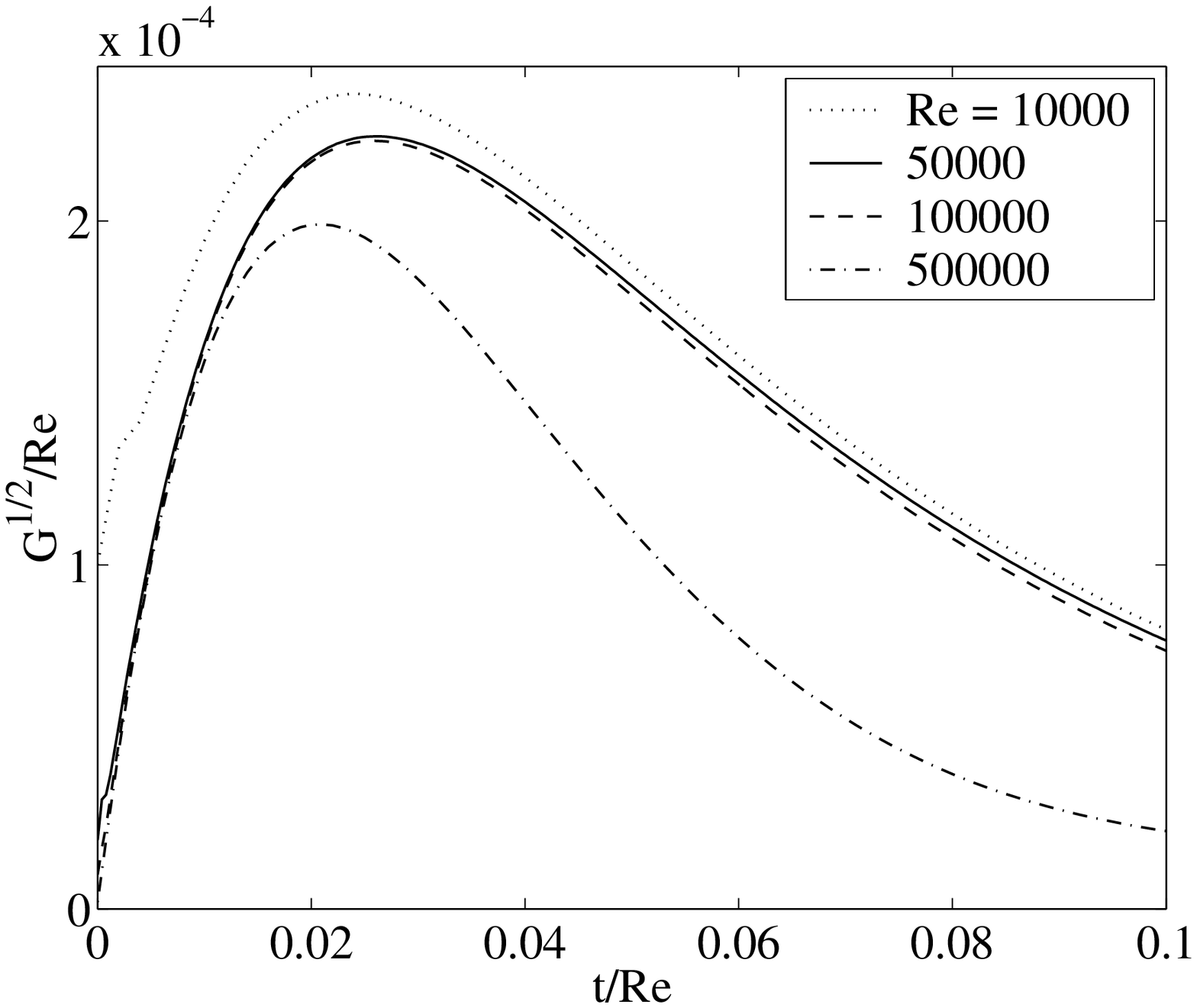,width=3.0in}}
\put(-3.1,2.1){(a)}
\end{picture}
\end{minipage}
\\
\begin{minipage}[t]{3.0in}
\begin{picture}(3.0,2.5)
\centerline{\psfig{figure=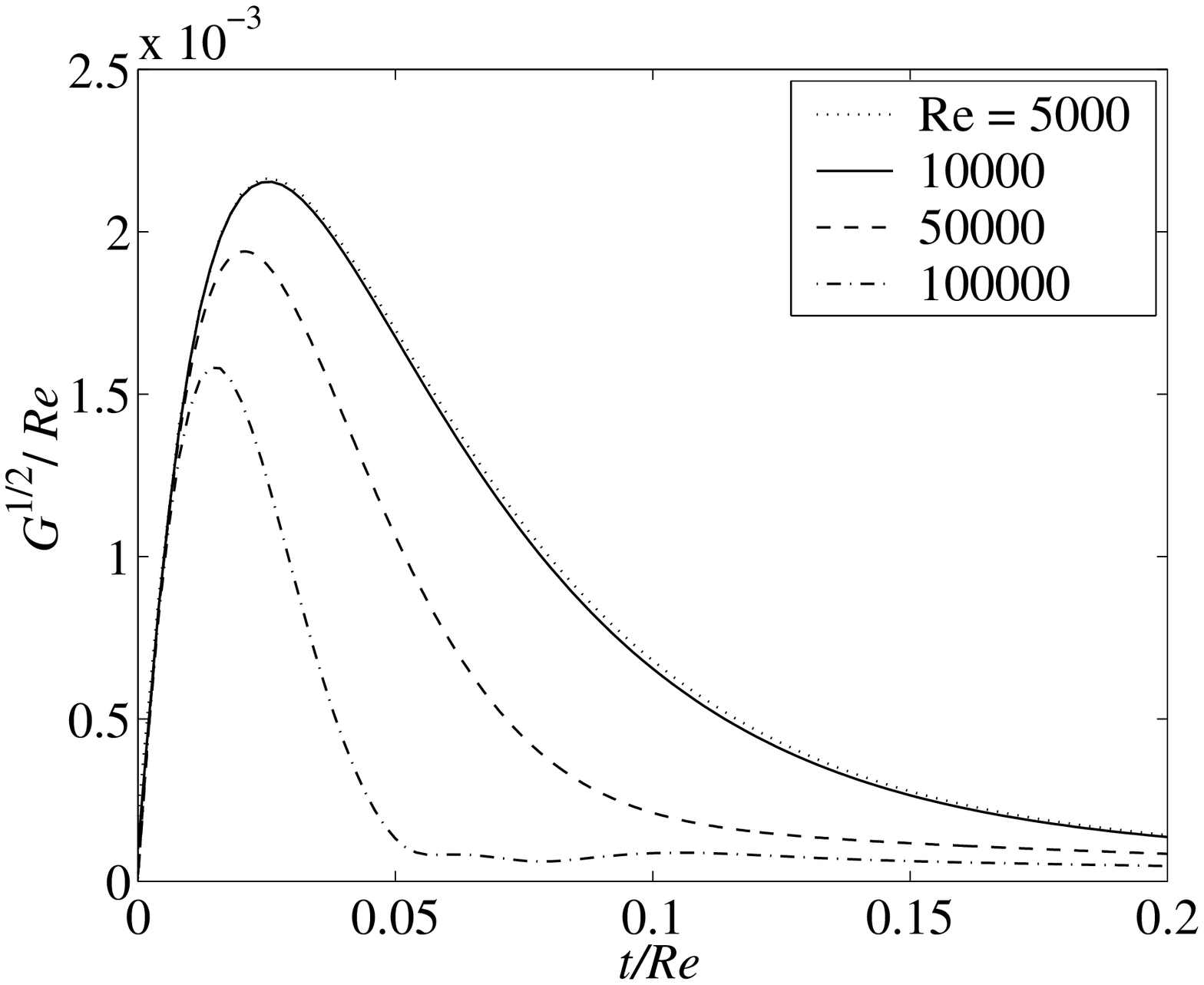,width=3.0in}}
\put(-3.1,2.1){(b)}
\end{picture}
\end{minipage}
\end{tabular}
\end{center}
\caption{Verification of the  quadratic scaling law for $G(t)$ for
streamwise-independent ($\alpha = 0$) disturbances.
(a) $\beta = 0.1$ and  $M = 2$;
(b) $\beta = 1.0$ and  $M = 2$.
}
\label{GR2tRal0bt01M2}
\end{figure}

For incompressible channel flows with streamwise-independent modes (i.e., $\alpha = 0$), 
Gustavsson~\cite{Gust91} has shown  
that $G_{max}$ varies quadratically with the Reynolds number $Re$,
and $t_{max}$ varies linearly with $Re$.
More specifically, when the energy growth curve, such as the one in Fig. 4($a$),
is plotted in terms of  $G(t)/{\it Re}^2$ and  $t/Re$,
the renormalized growth curves for different $Re$  collapse 
onto a single `universal' curve. Following a similar analysis,
Hanifi and Henningson~\cite{HH98} found that this scaling law 
also holds for compressible boundary layers.
However, for the present Couette flow of non-uniform shear and non-isothermal fluid,
this scaling law does not hold as shown in Fig.~\ref{GR2tRal0bt01M2} 
which displays  plots of $G(t)/{\it Re}^2$ verses $t/Re$ for a wide range of ${\it Re}$
at $M=2$. Similar trends persist at other values of Mach number 
(not shown for brevity). 

%-------------------------
%-------------------------

\begin{figure}[p]
\begin{center}
\begin{tabular}{c}
\begin{minipage}[t]{3.0in}
\begin{picture}(3.0,2.5)
\centerline{\psfig{figure=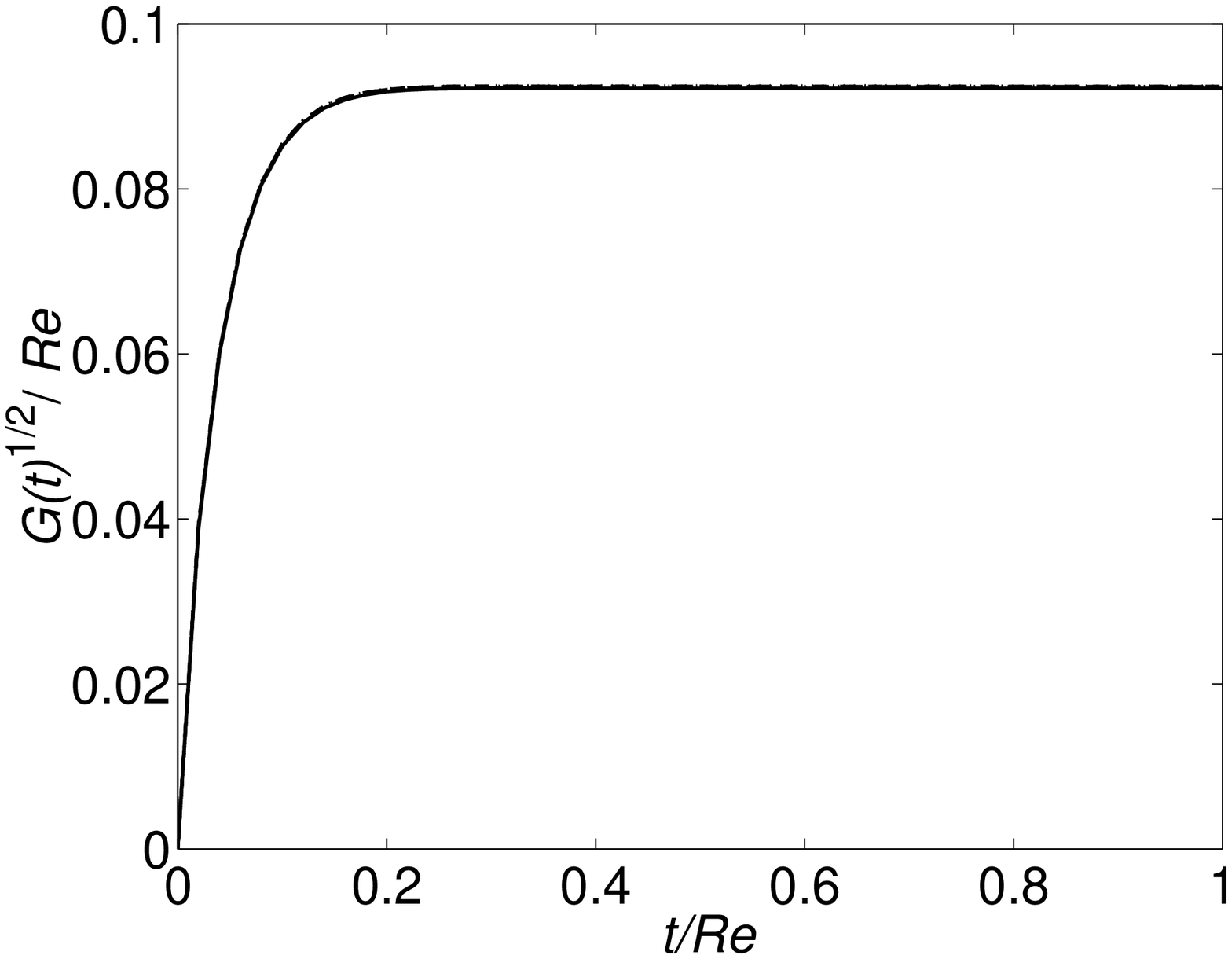,width=3.0in}}
\put(-3.1,2.1){(a)}
\end{picture}
\end{minipage}
\\
\begin{minipage}[t]{3.0in}
\begin{picture}(3.0,2.5)
\centerline{\psfig{figure=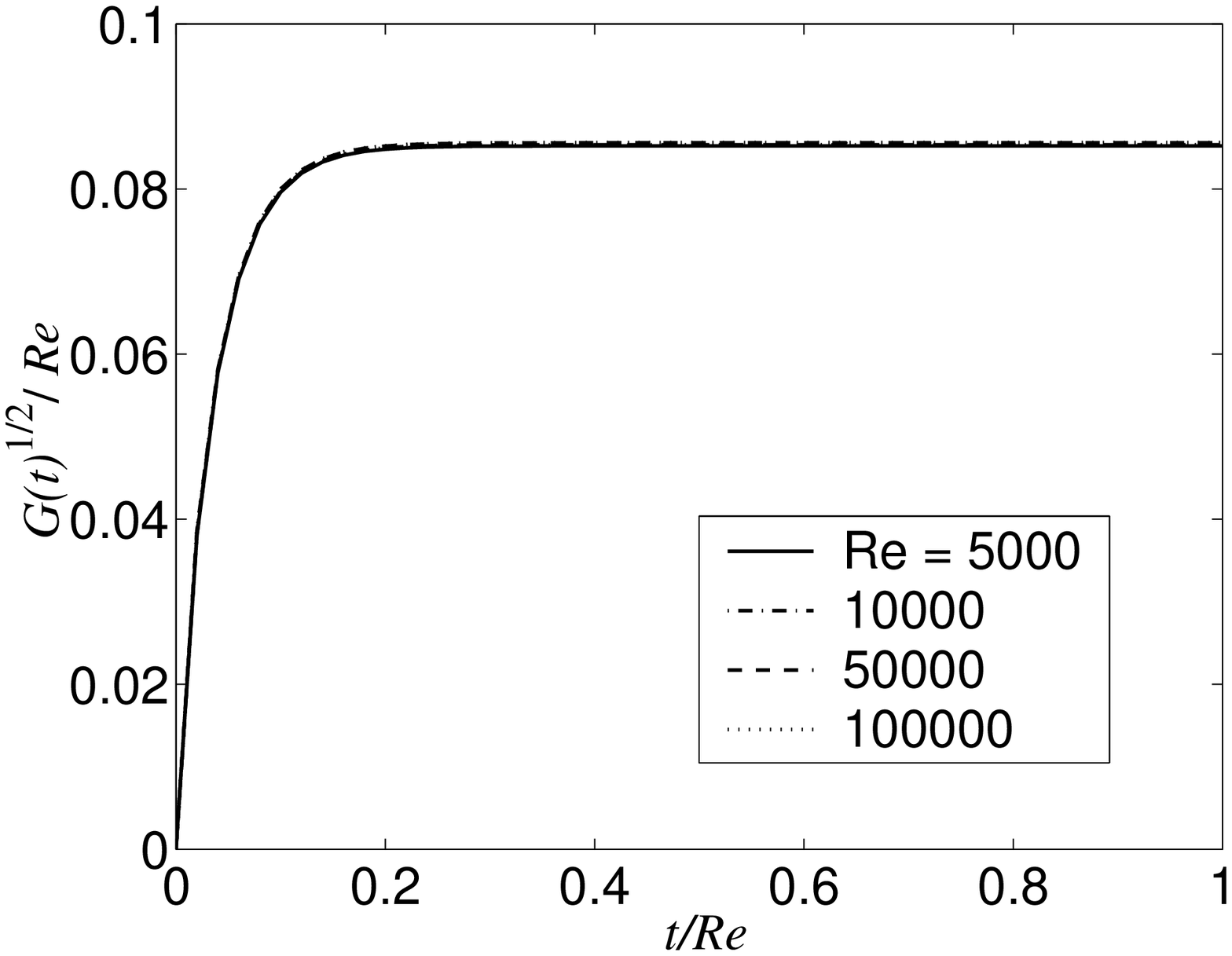,width=3.0in}}
\put(-3.1,2.1){(b)}
\end{picture}
\end{minipage}
\end{tabular}
\end{center}
\caption{Verification of the quadratic scaling law for $G(t)$ for
streamwise-independent ($\alpha = 0$) disturbances without 
pressure-related terms in ${\mathcal L}$ (see text for details). 
(a) $\beta = 0.1$ and  $M = 2$;
(b) $\beta = 1.0$ and  $M = 2$.
}
\label{fig:fig_Gscaling}
\end{figure}

For the present flow configuration, it can be verified that
the  Mack transformation~\cite{Mack84,HH98}
\begin{equation}
  \{ u', v', w', \rho', T', t\} 
   \rightarrow \{ {\it Re} \ \check{u}, \check{v}, \check{w}, {\it Re} \ 
       \check{\rho}, {\it Re} \ \check{T}, {\it Re} \ \check{t} \} 
\label{eqn_transformation1}
\end{equation}
(hence $p' \rightarrow {\it Re} \  \check{p}$, from the equation of state), 
can make the streamwise-independent linearized stability 
equation~(\ref{qstateeqn}) independent of 
the Reynolds number ${\it Re}$, except 
for the terms associated with density and temperature fluctuations
in the $y$- and $z$-momentum equations (i.e.,  
${\cal L}_{24}, {\cal L}_{25}, {\cal L}_{34}$ and ${\cal L}_{35}$ 
as detailed in the  Appendix). 
Only on neglecting these terms, one can show that $G_{\max} \sim {\it Re}^2$ as detailed
by Hanifi and Henningson~\cite{HH98} who further assumed that
$p' \rightarrow \check{p}/{\it Re}$ to neglect 
${\cal L}_{24}, {\cal L}_{25}, {\cal L}_{34}$ and ${\cal L}_{35}$ for large $\it Re$. 
The above pressure-related terms may not be negligible
for all mean flows. This can be ascertained for the present flow configuration
if we recompute the energy growth by setting  
\begin{equation}
  {\cal L}_{24}= {\cal L}_{25}= {\cal L}_{34} ={\cal L}_{35} =0
\label{eqn_pressterms}
\end{equation}
in the linear operator ${\mathcal L}$.
Indeed the rescaled growth curves for different Re now collapse onto
a single curve as shown in Figs. \ref{fig:fig_Gscaling}($a$-$b$),
for the same parameter values as in Figs. \ref{GR2tRal0bt01M2}($a$-$b$).
(Note that in these plots the energy growth does not decay 
since the above procedure (\ref{eqn_pressterms})
introduces some `artificial' neutral modes in ${\mathcal L}$,
resulting in an asymptotic value for $G(t)$ at large times.)
Therefore, we  conclude that the non-negligible values of 
${\cal L}_{24}, {\cal L}_{25}, {\cal L}_{34}$ and ${\cal L}_{35}$
for the plane Couette flow are responsible for 
the invalidity of the scaling law $G_{max}\sim Re^2$.

%-------------------------
%-------------------------

We note in passing that Hanifi and Henningson's~\cite{HH98} assumption
of $p' \rightarrow \check{p}/{\it Re}$ 
leads to a change in the equation of state by introducing a factor of $\it Re$ into it.
The form-invariance of the equation of state is essential
since it has been used to remove $p'$ from the linear perturbation system. 
In general, the (streamwise-independent) linear operator $\cal L$ of compressible flows 
cannot be made free from the apearance of $\it Re$ (via the Mack transformation), and hence 
the scaling, $G_{max}\sim Re^2$, would not hold for all mean flows. 
For special cases, the terms 
${\cal L}_{24}, {\cal L}_{25}, {\cal L}_{34}$ and ${\cal L}_{35}$ might be negligible
(e.g. in compressible boundary layers~\cite{HH98} at high $Re$) and hence the scaling law
would hold there.

\section{Nonmodal Energy Budget}

We now consider the evolution equation for the perturbation energy
${\cal E}(\alpha,\beta,t)$.
Multiplying (\ref{qstateeqn}) by ${\bf \tilde{q}^{\dagger} {\mathcal M}}$ 
and adding the complex 
conjugate of the resulting equation to itself, we  obtain the evolution equation 
for ${\cal E}(\alpha,\beta,t)$ as
\begin{equation}
  \frac{\partial \cal E}{\partial t} = -i \int_{-1}^{1}{\bf \tilde{q}^{\dagger} 
    {\mathcal M} {\cal L} \tilde{q}} \ {\rm d}\xi + c.c.
  = \sum_{j=0}^4 \dot{\cal E}_j, 
\label{Eevolv} 
\end{equation}
with $c.c.$ being the  complex-conjugate term. 
In the above equation, the total perturbation energy
has been decomposed into several constituent energies, $\dot{\cal E}_j$:
\begin{eqnarray}
\dot{\cal E}_1  &=&  -\int_{0}^{1} \left[\rho_0 U_{0y}
  \tilde{u}^{\dagger}\tilde{v} + \frac{T_0 \rho_{0y}}{\rho_0 \gamma M^2}
  \tilde{\rho}^{\dagger}\tilde{v} \right. \nonumber \\
  & &  \left. +\frac{\rho_0 T_{0y}}{T_0 \gamma (\gamma-1) M^2}
  \tilde{T}^{\dagger}\tilde{v} \right] {\rm d} y + c.c.
\label{Edot1}
\end{eqnarray}
\begin{eqnarray}
\dot{\cal E}_2 & = & -\frac{1}{\it Re}\int_{0}^{1} \left[\alpha^2
  (\mu_0 + \lambda_0) \tilde{u}^{\dagger}\tilde{u} + \mu_0(\alpha^2 + \beta ^2)
  \tilde{u}^{\dagger}\tilde{u}
  \right . \nonumber \\&& \left .
  - \tilde{u}^{\dagger}(\mu_{0y}D + \mu_0D^2)\tilde{u}
  -{\rm i}\alpha\tilde{u}^{\dagger}(\mu_{0y}
  \right . \nonumber \\&& \left .
   + (\mu_0 + \lambda_0)D)\tilde{v}
  + \alpha \beta (\mu_0 + \lambda_0) \tilde{u}^{\dagger}\tilde{w}
  \right . \nonumber \\&& \left .
  -(U_{0yy}\mu_T + U_{0y} T_{0y}\mu_{TT}) \tilde{u}^{\dagger}\tilde{T}
  -U_{0y}\mu_T \tilde{u}^{\dagger} D\tilde{T}
  \right . \nonumber \\&& \left .
  -{\rm i}\alpha\tilde{v}^{\dagger}(\lambda_{0y} + (\mu_0 + \lambda_0)D)\tilde{u}
  \right . \nonumber \\&& \left .
  + \mu_0(\alpha^2 + \beta^2)\tilde{v}^{\dagger}\tilde{v}
  -\tilde{v}^{\dagger}((\lambda_{0y} + \mu_{0y})D
  \right . \nonumber \\&& \left .
  + (\lambda_0 + \mu_0)D^2 +\mu_{0y} D + \mu_0 D^2)\tilde{v}
  \right . \nonumber \\&& \left .
  - {\rm i}\beta(\lambda_0 + \mu_0) \tilde{v}^{\dagger}D \tilde{w}
  -{\rm i}\alpha U_{0y}\mu_T \tilde{v}^{\dagger} \tilde{T}
  \right . \nonumber \\&& \left .
  -{\rm i}\beta \lambda_{0y} \tilde{v}^{\dagger} \tilde{w}
  -{\rm i}\beta \mu_{0y} \tilde{w}^{\dagger} \tilde{v}
  \right . \nonumber \\&& \left .
  + \alpha \beta (\mu_0 + \lambda_0) \tilde{w}^{\dagger}\tilde{u}
  - {\rm i}\beta(\lambda_0 + \mu_0) \tilde{w}^{\dagger}D \tilde{v}
  \right . \nonumber \\&& \left .
  + (\mu_0(\alpha^2 + \beta^2) + \beta^2(\lambda_0 + \mu_0))\tilde{w}^{\dagger}\tilde{w}
  \right . \nonumber \\&& \left .
  -\mu_0\tilde{w}^{\dagger}D^2\tilde{w} -\mu_{0y}\tilde{w}^{\dagger}D\tilde{w} \right ] {\rm d} y + c.c.
\label{Edot2}
\end{eqnarray}
\begin{eqnarray}
\dot{\cal E}_3 & = & \frac{1}{\sigma{\it Re}(\gamma -1) M^2}
   \int_{0}^{1} \rho_0\tilde{T}^{\dagger}\left[ \mu_T T_{0yy} + T_{0y}^2 \mu_{TT}
     \right. \nonumber \\
    &+& \left. 2T_{0y} \mu_T D -(\alpha^2+\beta^2)\mu_0
     + \mu_0 D^2 \right ]\tilde{T} {\rm d} y + c.c.
\label{Edot3}
\end{eqnarray}
\begin{eqnarray}
\dot{\cal E}_4  &=&  \frac{1}{Re}\int_{0}^{1} \rho_0 \left [
   2\mu_0 U_{0y}\tilde{T}^{\dagger} D \tilde{u} \right. \nonumber \\
   &+& \left.  2{\rm i}\alpha \mu_0 U_{oy}\tilde{T}^{\dagger}\tilde{v}
  + U_{0y}^2 \mu_T \tilde{T}^{\dagger}\tilde{T}
   \right ] {\rm d} y + c.c.
   \label{Edot5}
\end{eqnarray}
Here, $\dot{\cal E}_1$ represents the energy transfer from the mean  flow, 
$\dot{\cal E}_2$ is the viscous dissipation, 
$\dot{\cal E}_3$ the thermal diffusion and 
$\dot{\cal E}_4$ the shear-work,
respectively. 
Note that there is an additional term, $\dot{\cal E}_0$,
\begin{eqnarray}
 \dot{\cal E}_0  &=&  -i\alpha \int_{-1}^{1}U_0\left[\rho_0 
    \left( \tilde{u}^{\dagger}\tilde{u}
  + \tilde{v}^{\dagger}\tilde{v} + \tilde{w}^{\dagger}\tilde{w} \right) +\frac{T_0}
  {\rho_0 \gamma M^2} \tilde{\rho}^{\dagger}\tilde{\rho} 
   + \frac{\rho_0}
  {T_0 \gamma (\gamma -1) M^2} \tilde{T}^{\dagger}\tilde{T} \right] {\rm d}\xi + c.c.
  \nonumber \\
   &=& 0,
\label{Edot0}
\end{eqnarray}
representing the convective transfer of perturbation energy 
(by the mean flow), which is identically zero.
The decomposition of the total perturbation energy into different
constituents, as in (\ref{Edot1}-\ref{Edot5}), is useful to
analyse the role of each constituent energy on the growth/decay of total energy.
The evolution equation (\ref{Eevolv}) provides an energy-budget
for the total perturbation energy, and helps to quantify the contributions of  
different kinds of perturbation energies,
leading to the transient growth.

To analyse the nonmodal energy budget, we choose a set of values 
for $\alpha$, $\beta$, $M$ and ${\it Re}$ that leads to transient growth.
The constituent energies, $\dot{\mathcal E}_j$,
are then each evaluated using a quadrature formula with an initial perturbation 
configuration that would reach the optimal energy. 
The initial values of the constituent energies are chosen 
as $\{\dot{\cal E}(0)\} = \{1,0,0,0,0\}$ so that the total 
intitial energy is equal to the normalized value. 

%-------------------------
\begin{figure}[p]
\begin{center}
\begin{tabular}{c}
\begin{minipage}[t]{3.0in}
\begin{picture}(3.0,2.5)
\centerline{\psfig{figure=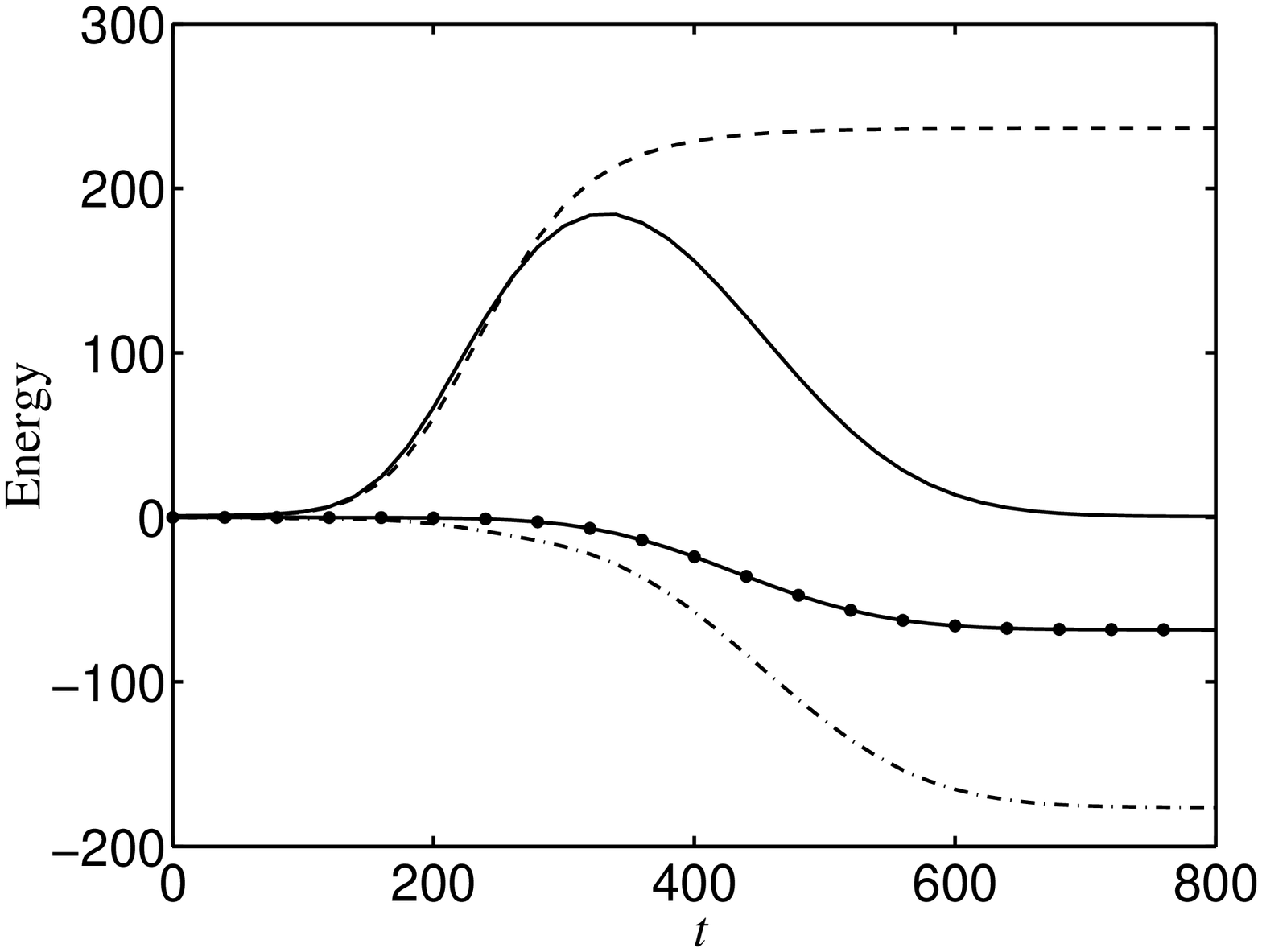,width=3.0in}}
\put(-3.1,2.1){(a)}
\end{picture}
\end{minipage}
\\
\begin{minipage}[t]{3.0in}
\begin{picture}(3.0,2.5)
\centerline{\psfig{figure=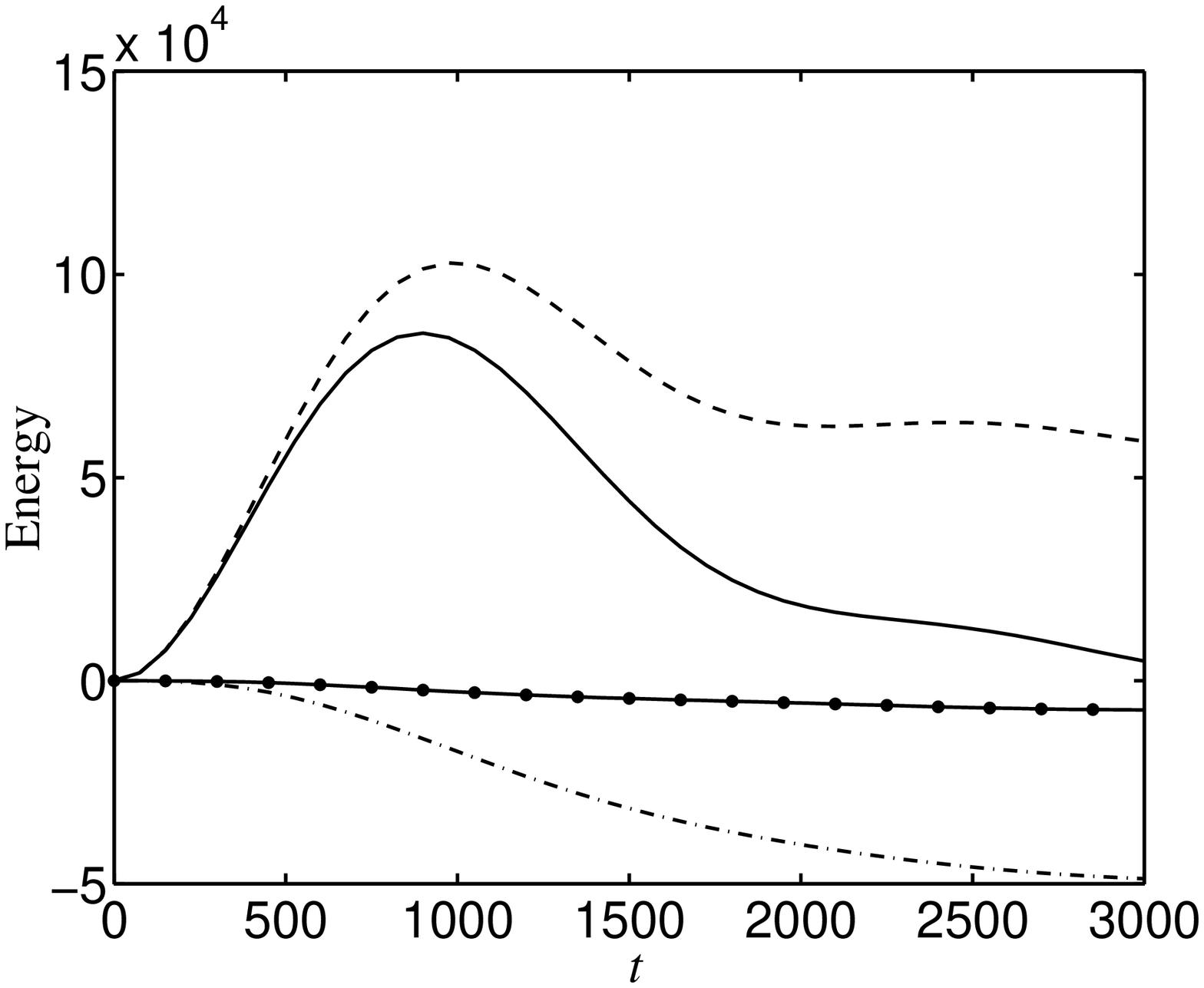,width=3.0in}}
\put(-3.1,2.1){(b)}
\end{picture}
\end{minipage}
\end{tabular}
\end{center}
\caption{Energy budget with time for $M = 2$ and ${\it Re} = 2 \times 10^5$; 
solid line, total energy; 
dash line, transferred from mean flow;  
dot-dash line, lost viscous dissipation; marked line, lost by thermal diffusion. 
(a) $\alpha = 0.1$, $\beta = 0.1$. (b) $\alpha = 0$ and $\beta = 3$}
\label{ebudgetR2e5M2al0105bt010}
\end{figure}
%-------------------------

Figure~\ref{ebudgetR2e5M2al0105bt010}($a$) shows the 
energy budget for three-dimensional disturbances with $\alpha=0.1=\beta$;
other parameters are as in Fig. 4($a$).
For this case, the energy gain  by the shear-work term, $\dot{\mathcal E}_4$,
is found to be negligible and hence not shown in this plot. 
It is observed  that the energy 
transferred from the mean flow, $\dot{\mathcal E}_1$, (denoted by the dash line)
increases with time, reaching a large asymptotic value beyond the optimal time ($=t_{max}$);
this is the major cause for the transient growth.  
At some later time ($t>t_{max}$), this transient growth is nullified by
the energy taken away by the thermal diffusion, 
$\dot{\mathcal E}_3$, (denoted by the  line with solid symbols) 
and the viscous dissipation, $\dot{\mathcal E}_2$ (dot-dash line). 
For this parameter set, the energy loss due to the viscous dissipation
dominates over that due to the thermal diffusion at large times.

Figure ~\ref{ebudgetR2e5M2al0105bt010}($b$) shows the
same energy budget for a representative case 
of streamwise independent ($\alpha=0$) disturbances,
with parameter values as in  Fig. 8.
As in the previous case (Fig. \ref{ebudgetR2e5M2al0105bt010}$a$), the energy 
transferred from the mean flow is responsible for the transient growth;
however, unlike in Fig.~\ref{ebudgetR2e5M2al0105bt010}($a$), $\dot{\mathcal E}_1$
reaches a peak  and then decreases to attain an asymptotic value.  
The variations of the other constituent energies are similar
to those in Fig.~\ref{ebudgetR2e5M2al0105bt010}($a$).

%-------------------------
\begin{figure}[p]
\begin{center}
\begin{tabular}{c}
\begin{minipage}[t]{3.0in}
\begin{picture}(3.0,2.5)
\centerline{\psfig{figure=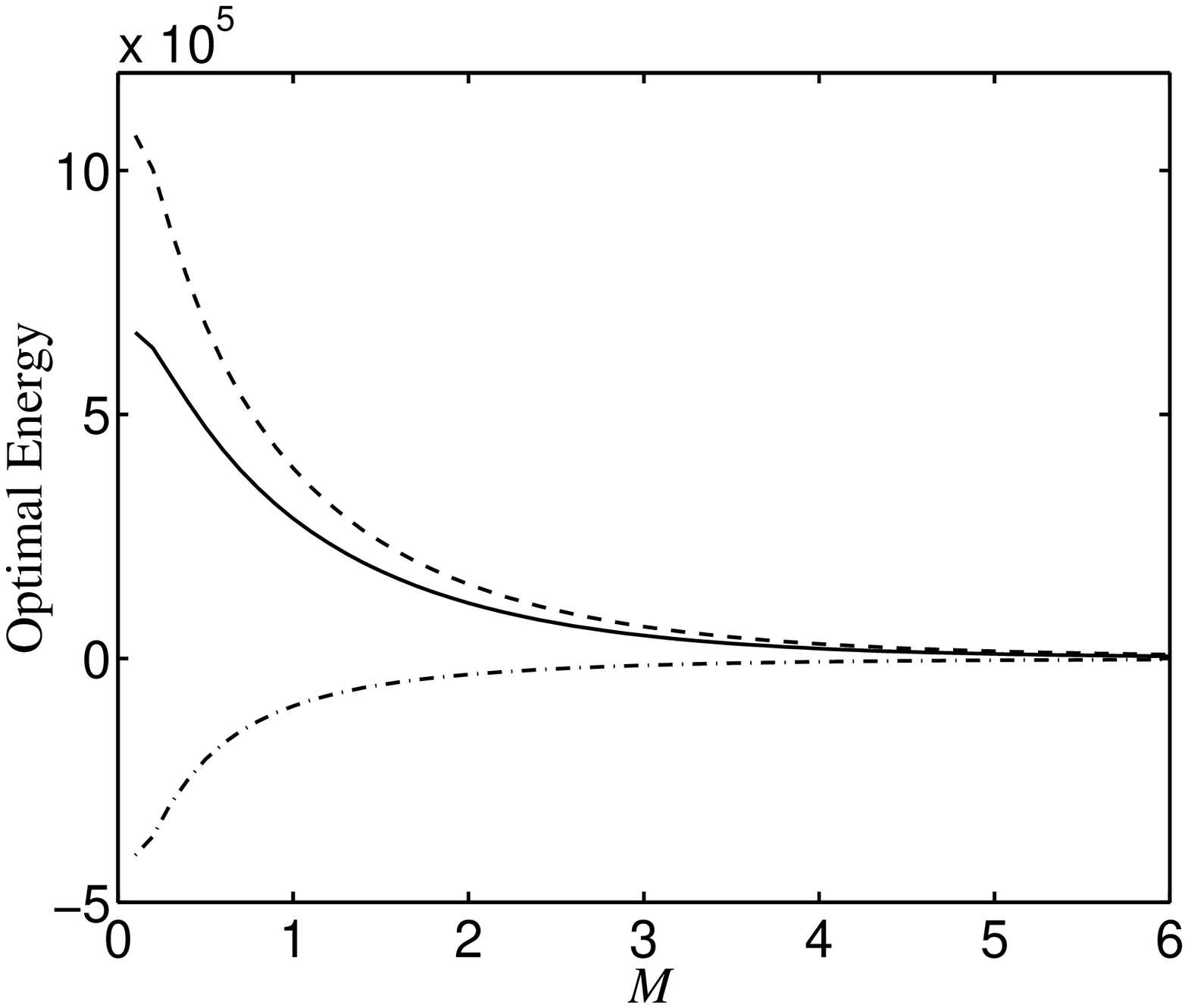,width=3.0in}}
\put(-3.1,2.1){(a)}
\end{picture}
\end{minipage}
\\
\begin{minipage}[t]{3.0in}
\begin{picture}(3.0,2.5)
\centerline{\psfig{figure=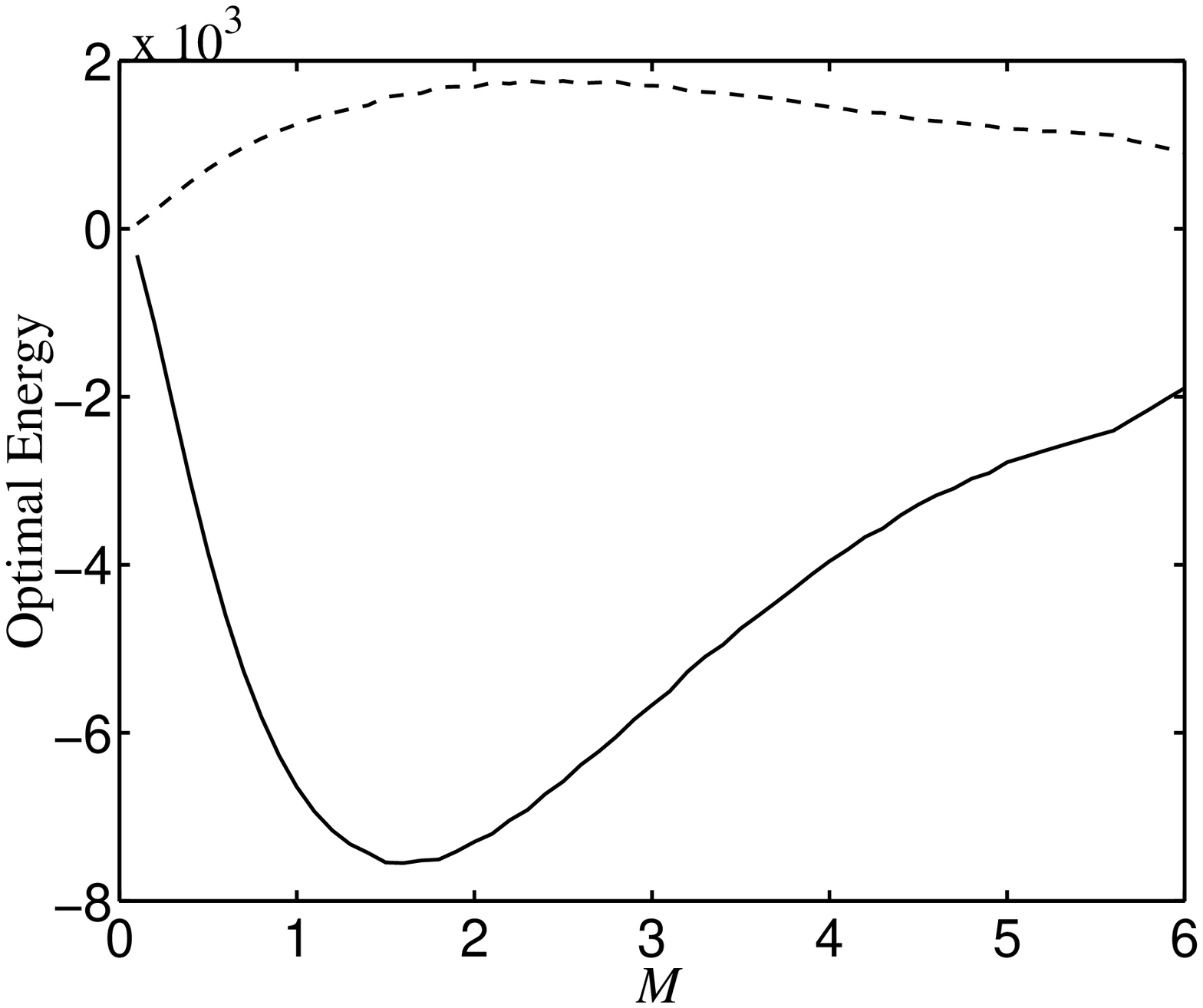,width=3.0in}}
\put(-3.1,2.1){(b)}
\end{picture}
\end{minipage}
\end{tabular}
\end{center}
\caption{Energy budget with  Mach number at $(\alpha_{\rm opt},\beta_{\rm opt},t_{\rm opt})$ 
for ${\it Re} = 10^5$. (a) solid line, total energy; dash line, transferred from mean flow; 
dot-dash line, lost viscous dissipation. (b) solid line, energy lost by thermal diffusion; 
dash line, energy gained by the shear-work term.}
\label{optebudgetR1e5}
\end{figure}
%-------------------------

In the previous section, we found that the optimal transient
energy growth $G_{opt}$ decreases with increasing Mach number.
To explain this behaviour in terms of constituent energies,
we show the energy budget versus Mach number in 
Fig.~\ref{optebudgetR1e5} for the optimal 
values of $(\alpha,\beta,t)$ (as in Fig. ~\ref{GtalbtoptxMRe1e5}). 
Note that the total energy curve   
shown in Fig.~\ref{optebudgetR1e5}($a$) (denoted by the solid line) matches with the one 
in Fig.~\ref{GtalbtoptxMRe1e5}($a$). 
It is observed  that the energy transferred from the mean 
flow ($\dot{\cal E}_1$) and the  viscous dissipation loss ($\dot{\cal E}_2$) are 
the most significant compared to  the other components ($\dot{\cal E}_3$
and $\dot{\cal E}_4$, see Fig.~\ref{optebudgetR1e5}$b$) of the perturbation energy.
Since $\dot{\cal E}_1$ decreases with increasing $M$, so does 
the total energy at the optimal time.
In general, the decrease in the transient energy growth with increasing $M$ is
primarily due to the decrease of the energy transfer from the
mean flow to perturbations in the same limit.

For the present flow configuration, it is  still not clear why the transferred
energy from the mean flow, $\dot{\cal E}_1$, decreases with increasing  Mach number;
an inviscid energy analysis provides an answer for this
as well as for the  transient growth mechanism.

\subsection{Inviscid limit: Mechanism for transient growth}

For streamwise independent disturbances ($\alpha=0$),
Hanifi and Henningson~\cite{HH98} 
have shown that the inviscid linear stability equations 
of compressible fluids have the following solution
\begin{equation}
   {\bf \tilde{q}}_{\rm ivs}(y,t) = \hat{A}{\bf \tilde{q}}_{\rm ivs}(y,0), 
\label{eqn_invis_sol} 
\end{equation}
where
\begin{equation}
  \hat{A} = {\rm I} + \tilde{A} \quad \mbox{and} \quad
  {\bf \tilde{q}}_{\rm ivs}(y,t) = \{\tilde{u}_{\rm ivs},\tilde{v}_{\rm ivs},
  \tilde{\rho}_{\rm ivs},\tilde{T}_{\rm ivs}\}^{\rm T},
\end{equation}
and the subscript, `ivs' stands for `inviscid'. 
The nonzero elements of the matrix, $\tilde{A}$, are $\tilde{A}_{12} = -U_{0y}t$, 
$\tilde{A}_{32}=-\rho_{0y}t$ and $\tilde{A}_{42}=-T_{0y}t$.  
In fact, this inviscid solution for velocity perturbations
is the same as that of Ellingsen and Palm's incompressible solution~\cite{EP75}.
The inviscid perturbation energy, $G_{\rm ivs}(t)$, 
maximized over $\tilde{q}_{\rm ivs}(y,0)$, in the same definition of the 
energy norm, is  given by
\begin{equation}
   G_{\rm ivs}(t) = \max(\{\hat{\lambda}_k\}),
\end{equation}
where $\{\hat{\lambda}_k\}$'s are the eigenvalues of the differential equation
\begin{equation}
  \hat{A}^{\dagger}\hat{\mathcal L}\hat{A}\tilde{q}_{\rm ivs}(y,0) 
   = \hat{\lambda}\hat{\mathcal L}\tilde{q}_{\rm ivs}(y,0),
\end{equation}
with $\hat{\mathcal L}$ being  the associated linear differential operator
\begin{equation}
   \hat{\mathcal L}=\mbox{diag}\{\rho_0,\rho_0(1-\beta^{-2}{\rm d}/{\rm d}y),T_0^2/\gamma M^2,
     \rho_0^2/\gamma(\gamma-1)M^2\}.
\end{equation}
The above equation is solved using the same spectral method described in Sec. 3.1
with the following boundary conditions:
\begin{equation}
  \tilde{v}_{\rm ivs}(0,0) = \tilde{v}_{\rm ivs}(1,0) = 0 .
\end{equation}

%-------------------------
\begin{figure}[p]
\begin{center}
\begin{tabular}{c}
\begin{minipage}[t]{3.0in}
\begin{picture}(3.0,2.5)
\centerline{\psfig{figure=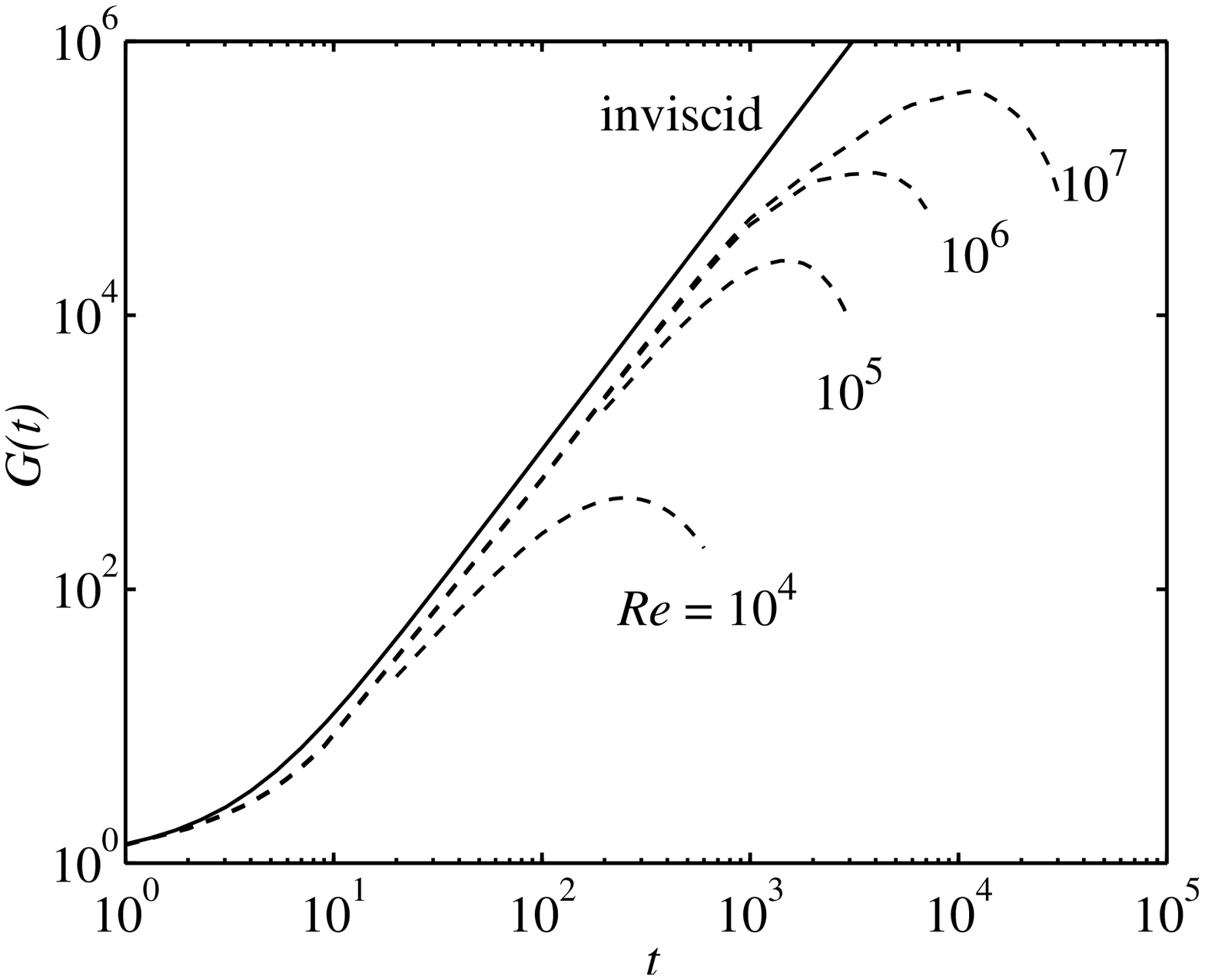,width=3.0in}}
\put(-2.625,1.4){\psfig{figure=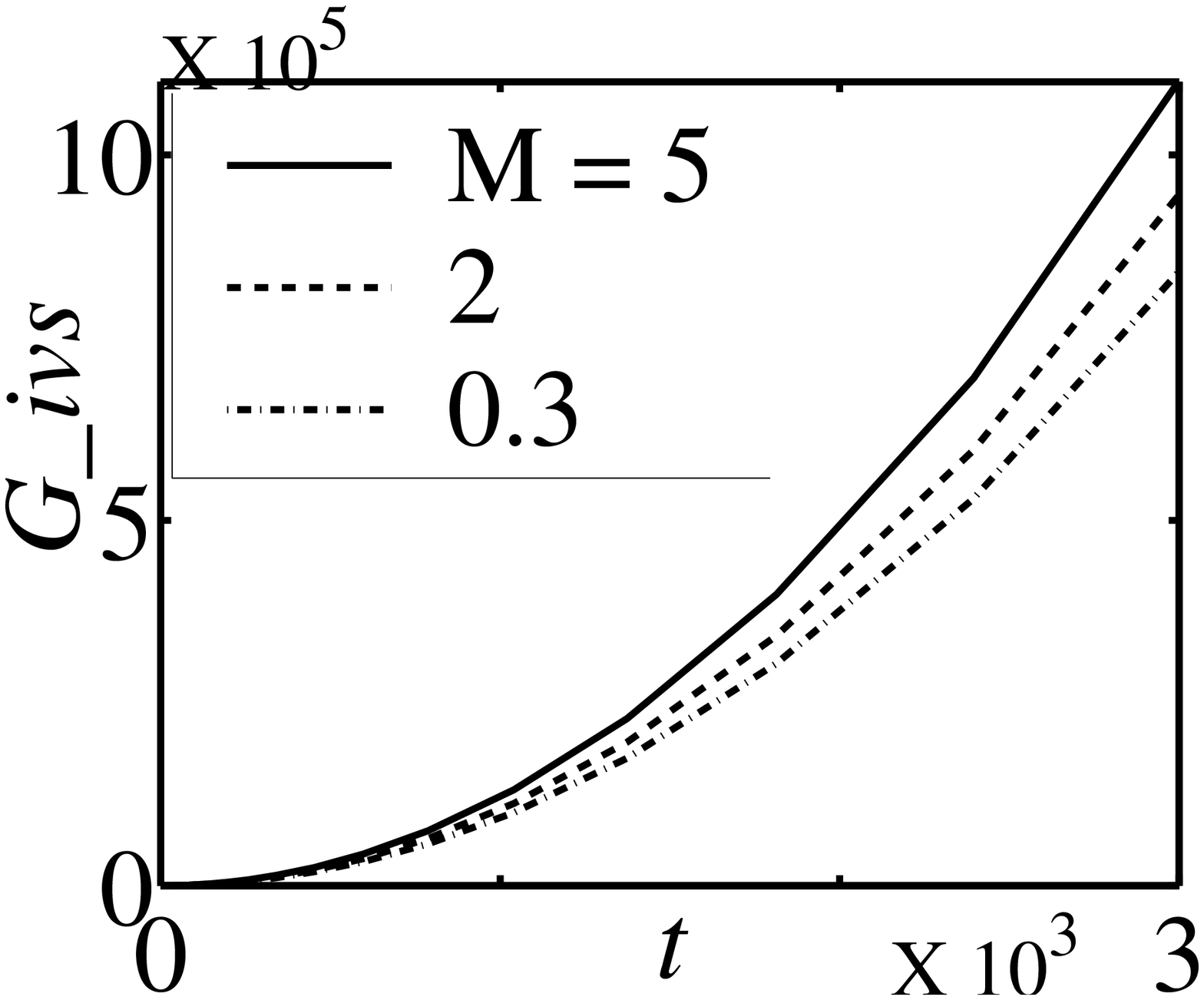,width=1.05in}}
\end{picture}
\end{minipage}
\end{tabular}
\end{center}
\caption{Inviscid energy growth ($G_{ivs}(t)$) and its comparison with 
viscous growth curves at various Reynolds number
for $M = 2$, $\alpha = 0$ and $\beta = 1$.
The inset shows the variations of $G_{ivs}(t)$ for three
different Mach numbers.
}
\label{Gtinvisal0bt1}
\end{figure}
%-------------------------

Figure~\ref{Gtinvisal0bt1}
shows the variation of $G_{\rm ivs}(t)$ with time for $\beta = 1$ and $M = 2$.
For comparison, the viscous $G(t)$-curves at various $\it Re$ are also displayed. 
In terms of different energy components, the inviscid energy growth rate of the optimal 
perturbation configuration is given by (\ref{Edot1}),
since the transfer rates of the other energy components, $\dot{\cal E}_2,\ \dot{\cal E}_3$ 
and $\dot{\cal E}_4$ are zero in the inviscid limit.
Therefore the reason for the inviscid growth is due to the 
energy transferred from the mean flow to perturbations, i.e. $\dot{\cal E}_1$. 
This energy transfer occurs due to an {\it algebraic 
instability}~\cite{Land80,EP75},
wherein the streamwise perturbation velocity, density and temperature
grow algebraically (linearly) with time.
if the initial normal perturbation velocity is non-zero 
(i.e. $\tilde{v}_{ivs}(0)\neq 0$).
A noteworthy point in Fig.~\ref{Gtinvisal0bt1} that the 
inviscid growth-curve, $G_{\rm ivs}(t)$, 
coincides with the viscous curves only {\it for a very short time}--
we shall return to explain this point later.  Note that
this observation is in contrast to boundary layers (see Figs. 1 and 2 in ref.~\cite{HH98})
for which the  viscous growth curves (at different $Re$)
closely follow the inviscid growth-curve till they achieve their maxima $G_{max}$.

The inset in Fig.~\ref{Gtinvisal0bt1} shows the curves 
of $G_{ivs}(t)$ ($=\dot{\mathcal E}_1$)
for three different Mach numbers. Clearly, at any time,
$\dot{\mathcal E}_1$ increases with increasing $M$ for the inviscid case.
This is in contrast to the viscous case where we have
seen a steady decrease of $\dot{\mathcal E}_1$ with increasing $M$ 
(dashed line in Fig. 15$a$).
To understand this difference,
let us rewrite the energy transfer rate (\ref{Edot1}) as
\begin{equation}
  \dot{{\cal E}}_1(t) = -\int_0^1 {\bf \tilde{q}}^{\dagger}{\mathcal M}
     {\bf q_{0y}}\tilde{v}\: dy 
          + c.c. ,
\label{Edot1_version}
\end{equation}
which is identical for both the inviscid and viscous cases.
Now combining (\ref{Edot1_version}) 
with (\ref{eqn_invis_sol}), we obtain 
\begin{equation}
  \dot{{\cal E}}_{\rm ivs}(t) = -2\int_0^1 {\bf \tilde{q}}_{\rm ivs}^{\dagger}(y,0)
      \hat{A}^{\dagger}\hat{\mathcal L}{\bf q_{0y}}\tilde{v}_{\rm ivs}(y,0) \: dy
\label{Edot1_invis}
\end{equation}
In this equation the $M$-dependence is present only via
$\hat{A}^{\dagger}\hat{\mathcal L}{\bf q_{0y}}$.
Note that the inviscid temperature eigenfunction,
\begin{equation}
 \tilde{T}_{ivs}(t) = \tilde{T}_{ivs}(0) - \tilde{v}_{ivs}(0) T_{0y} t,
\end{equation}
grows quadratically with $M$ since $T_{0y}\sim M^2$.
It can be verified that the dominant contribution to $\dot{{\cal E}}_{\rm ivs}(t)$
in  (\ref{Edot1_invis})
at large $M$ comes from the energy associated with temperature fluctuations,
and hence, for any given initial condition, the norm of $\dot{{\cal E}}_{\rm ivs}(t)$
would increase with increasing $M$.
We can conclude that the increase of the inviscid energy growth with increasing $M$ 
is primarily due to the  increased energy transfer from the mean 
flow to {\it temperature fluctuations}.

Though the above contribution is also present for the {\it viscous} case,
the decay of the viscous eigenfunction 
${\bf \tilde{q}}$ by viscosity $\mu$ becomes dominant with increasing $M$
as we show below.
Compared to the inviscid eigenfunction ${\bf \tilde{q}}_{\rm ivs}$,
the viscous eigenfunction ${\bf \tilde{q}}$  decays due to  
viscous dissipation and  thermal diffusion (both of which involve  viscosity $\mu$). 
Under the assumed viscosity-law (\ref{eqn_viscosity}), the viscosity of the mean flow
increases rapidly with increasing Mach number at all points in the normal direction
(see Fig. 1), resulting in a decrease in the viscous eigenfunction ${\bf \tilde{q}}$
in the same limit. To quantify  the last statement, we calculate the following 
for the viscous problem:
\begin{equation}
   {\cal E_{\rm ef}}=\frac{\sum_l\int_0^1 {\rm Tr} \ \tilde{q}^{(l)\dagger}(t,y)
      \tilde{q}^{(l)}(t,y) {\rm d}y}{\sum_l\int_0^1 {\rm Tr} \ 
      \tilde{q}^{(l)\dagger}(0,y)\tilde{q}^{(l)}(0,y) {\rm d}y} 
       = \frac{\sum_{l,k}|\kappa_k(0)|^2e^{2\;{\rm Im}(\omega_k)t} 
          \int_0^1 |q'^{(l)}_k(y)|^2 {\rm d}y}{\sum_{l,k}|\kappa_k(0)|^2 
          \int_0^1 |q'^{(l)}_k(y)|^2 {\rm d}y},
\label{eqn_ef}
\end{equation}
which is a measure of the {\it collective evolution} (growth/decay) of all eigenfunctions.
Here, the summation index $l$ runs over the perturbation velocity, density and temperature, 
and the index $k$ runs over the selected eigenmodes (see eqn. 19);
${\rm Im}$ stands for the imaginary part, 
and ${\rm Tr}$ stands for the trace of the matrix. 
The initial condition, $\{\kappa(0)\}$, is chosen as $\{1\}$.
It should be pointed out that the measure for the collective
growth/decay of eigenfunctions, via  (\ref{eqn_ef}), is equivalent to probing the
energy-norm with an unit weight matrix ${\mathcal M} =I$.
Our definition simply masks out the Mach-number dependence of the energy norm
due to the base-state variables in ${\mathcal M}$ (as in the Mack energy norm).

%---------------------------
\begin{figure}
\centerline{\psfig{figure=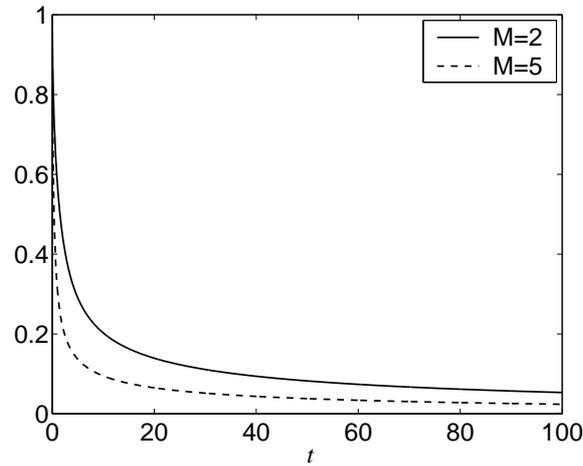,width=3in}}
\caption{Variation of the collective growth/decay of all eigenfunctions,
${\cal E_{\rm ef}}$, with time: 
$\alpha = 0$, $\beta = 1$ and ${\it Re} = 2\times 10^5$}.
\label{fig:fig_eigenfunction1}
\end{figure}
%---------------------------

Figure \ref{fig:fig_eigenfunction1} shows the variation of
${\cal E_{\rm ef}}$ with time for two Mach numbers;
other parameters are $\alpha=0$ and $\beta=1$.
It is clear from this figure that the (collective) decay-rate of eigenfunctions 
at any time is higher for higher Mach number.
Therefore, the viscous eigenfunctions ${\bf \tilde{q}}$ are small at a time
$t=t_{\rm opt}$ or $t_{\max}$ and at a Mach number $M_0$ in comparison with 
that at any lower value of the Mach number $M < M_0$, leading to a decrease in
the energy transfer from the mean flow to perturbations with increasing Mach number. 
This decay can only come from the dependence of the viscous eigenfunctions
on the shear viscosity (which increases with increasing $M$).
In contrast, the inviscid eigenfunctions grow with $M$ as we have pointed out earlier.
Therefore,  {\it the difference in the variation of $\dot{\mathcal E}_1$
with $M$ for the inviscid and viscous cases
stems from the different variations of the respective eigenfunctions with $M$}.

The reason for the viscous growth curves (in Fig. 16) 
not coinciding with the inviscid curve 
during the  most part of the transient growth  
can be related to the fact  that the viscous eigenfunctions undergo 
sharp decays with time for the present mean flow.
(Note that the non-dimensional viscosity $\mu_0$ takes a value that is
always greater than one in the entire domain, see Fig. 1$d$.)
Here, the viscous-decay of eigenfunctions with time (as in Fig. 17)
competes with the inviscid growth of eigenfunctions.
Clearly, the decay of $\dot{\mathcal E}_1$
with time is not negligible compared to the algebraic growth 
even duing the initial times, and
this decrease becomes more and more important at later times.
Hence, the $G(t)$-curves for various $Re$ do not coincides with
their inviscid counterpart $G_{ivs}(t)$.

Returning to the compressible boundary layers~\cite{HH98},
we had noted that the viscous growth curves, $G(t)$, coincide 
with the inviscid curve $G_{\rm ivs}(t)$ till they achieve their maxima $G_{max}$.
A plausible explanation for this could be that
the temperature, and hence the viscosity, for boundary layers remains constant for 
more than $99\%$ of the domain 
(since the variations of mean-fields are concentrated within a few displacement thickness
while the mean flow extends upto a few hundred times of the displacement thickness).
Therefore, the viscous-decay of ${\bf q}$, compared 
to the inviscid (algebraic) growth, may not be as strong in
boundary layers as in the present mean flow.
This issue needs further investigation.

\section{Summary and Conclusion}

We have investigated the nonmodal transient growth characteristics and 
the related patterns in  compressible plane Couette flow 
of a perfect gas with temperature-dependent viscosity.
The mean flow consists of a non-uniform shear-field and
varying temperature and viscosity along the wall-normal direction. 
For the  transient growth analysis, the disturbance size was measured
in terms of the Mack energy norm~\cite{Mack69,Mack84}
for which the pressure-related energy transfers are zero.
The results were presented for  ranges of Mach number ($M$), Reynolds number ($Re$)
and wavenumbers ($\alpha$ and $\beta$) for which the flow is asymptotically stable.

The maximum transient energy growth, $G_{\max}$, is found to increase with  
increasing Reynolds number, as in many incompressible flows,
but decreases with increasing Mach number.
The optimal energy growth, $G_{\rm opt}$, (i.e. the global maximum of $G_{\max}$ 
in the $(\alpha, \beta)$-plane for  given $Re$ and $M$) decreases
with increasing $M$. 
This result is in  contrast to that for compressible
boundary layers~\cite{HSH96} for which
$G_{\rm opt}$ increases with increasing $M$.
The optimal streamwise wavenumber, $\alpha_{\rm opt}$, is close to zero (but finite)
at $M\to 0$, increases with increasing $M$ and reaches a maximum value at some
value of $M$, and decreases thereafter. 
Unlike in incompressible Couette flow~\cite{BF92,SH01}, $\alpha_{\rm opt}$
becomes  {\it zero} at large enough value of $M$.
The optimal spanwise wavenumber, $\beta_{\rm opt}$, also varies non-monotonically
with $M$: $\beta_{\rm opt}$ decreases first and then increases with increasing $M$.
Optimal velocity patterns (at $t=t_{max}$) correspond to {\it pure streamwise vortices}
for large $M$, but the modulated streamwise vortices
are optimal patterns for low-to-moderate values of $M$.
Our result on optimal patterns at very high Mach number 
should be contrasted with that for the incompressible Couette flow
for which the {\it oblique modes} constitute the  optimal patterns.

For the streamwise independent disturbances ($\alpha=0$),
we have found that the transient energy growth does not follow the well-known
scaling laws, $G_{\max} \sim {\it Re}^2$ and $t_{\max} \sim  {\it Re}$,
of incompressible shear flows~\cite{Gust91,SH01}.
In contrast, however, these scaling laws are known
to hold for compressible boundary layers~\cite{HSH96}. 
We showed that the invalidity of these scaling laws 
for the present flow configuration is tied   
to the `dominance' of some terms (related to
density and temperature fluctuations in the $y$ and $z$-momentum equations) 
in the linear stability operator. 
More specifically, we found that the well-known Mack transformation
(eqn.~\ref{eqn_transformation1}) does not make the streamwise-independent stability equations
independent of the Reynolds number because of the above mentioned dominant terms.

An evolution equation for the perturbation energy
has been derived, and various constituent energies, that
are transferred to perturbations through different physical processes, have been identified.
We have carried out  a detailed nonmodal energy  analysis
for initial perturbations that yield
maximum energy growth at a later time.
Based on this energy budget analysis,
we  found  that the  transient energy growth occurs due to
the transfer of energy from the mean flow to perturbations via an inviscid 
{\it algebraic} instability.
We further  showed that the decrease of transient growth 
with increasing Mach number is  tied to 
the decrease in the energy transferred from the mean flow 
(${\mathcal E}_1$) in the same limit.
Lastly, considering the inviscid limit of stability equations,
we found that the viscous growth curves follow the
inviscid growth curve ($G_{ivs}$) only for a very short time.
This is due to the strong dependence of viscosity on Mach number
for the present mean flow, resulting in  sharp decays of 
the viscous eigenfunctions with increasing  Mach number
which is responsible for the decrease of $\dot{\mathcal E}_1$
in the same limit.

%-------------------------

\end{document}